\documentclass[fleqn,usenatbib]{mnras}

\usepackage{newtxtext,newtxmath}
\usepackage[T1]{fontenc}

\DeclareRobustCommand{\VAN}[3]{#2}
\let\VANthebibliography\thebibliography
\def\thebibliography{\DeclareRobustCommand{\VAN}[3]{##3}\VANthebibliography}

\usepackage{graphicx}	% Including figure files
\usepackage{amsmath}	% Advanced maths commands
\usepackage{subcaption}
\usepackage{threeparttable}
\usepackage{multirow}
\newcommand{\unsim}{\mathord{\sim}}

\newcommand{\msun}{M$_\odot$}
\newcommand{\rsun}{R$_\odot$}
\newcommand{\kms}{$\mathrm{km}\,\mathrm{s}^{-1}$}

\title[The wTTS LkCa 4 observed with SPIRou and TESS]{The active weak-line T Tauri star LkCa 4 observed with SPIRou and TESS}

\author[B. Finociety et al.]{B. Finociety$^{1}$ \thanks{E-mail: benjamin.finociety@irap.omp.eu},
J.-F. Donati$^{1}$, K. Grankin$^{2}$, J. Bouvier$^{3}$, S. Alencar$^{4}$, F. Ménard$^{3}$, T.P. Ray$^{5}$, \newauthor Á. Kóspál$^{6}$ and the SLS consortium
\\
$^{1}$ Université de Toulouse, CNRS, IRAP, 14 av. Belin, 31400 Toulouse, France \\
$^{2}$ Crimean Astrophysical Observatory, 298409 Nauchny, Republic of Crimea \\
$^{3}$ Université Grenoble Alpes, CNRS, IPAG, F-38000 Grenoble, France \\
$^{4}$ Departamento de Fisica - ICEx - UFMG, Av. Antonio Carlos 6627, 30270-901 Belo Horizonte, MG, Brazil \\
$^{5}$ Dublin Institute for Advanced Studies, Dublin, Ireland \\
$^{6}$ Konkoly Observatory, Research Centre for Astronomy and Earth Sciences, Konkoly-Thege Miklós út 15-17, 1121 Budapest, Hungary 
}

\date{Accepted 2023 January 23. Received 2023 January 23; in original form 2022 November 25}

\pubyear{2022}

\begin{document}
\label{firstpage}
\pagerange{\pageref{firstpage}--\pageref{lastpage}}
\maketitle

\begin{abstract}
We report results of a spectropolarimetric and photometric monitoring of the weak-line T~Tauri star LkCa~4 within the SPIRou Legacy Survey large programme, based on data collected with SPIRou at the Canada-France-Hawaii Telescope and the TESS space probe between October 2021 and January 2022. We applied Zeeman-Doppler Imaging to our spectropolarimetric and photometric data to recover a surface brightness distribution compatible with TESS photometry, as well as the large-scale magnetic topology of the star. As expected from the difference in wavelength between near-infrared and optical data, the recovered surface brightness distribution is less contrasted than the previously published one based on ESPaDOnS data, but still features mid-latitude dark and bright spots. The large-scale magnetic field is consistent in shape and strength with the one derived previously, with a poloidal component resembling a 2.2~kG dipole and a toroidal component reaching 1.4 kG and encircling the star at the equator. Our new data confirm that the surface differential rotation of LkCa~4 is about 10 times weaker than that of the Sun, and significantly different from zero. Using our brightness reconstruction and Gaussian Process Regression, we were able to filter the radial velocity activity jitter down to a precision of $0.45$ and $0.38$~\kms\ (from an amplitude of 6.10~\kms), respectively, yielding again no evidence for a close-in massive planet orbiting the star.
\end{abstract}


%%%%%%%%%%%%%%%%%%%%%%%%%%%%%%%%%%%%%%%%%%%%%%%%%%

%%%%%%%%%%%%%%%%% BODY OF PAPER %%%%%%%%%%%%%%%%%%

\section{Introduction}

At an age of a few Myr, young low-mass stars (M$_*<2$~\msun) emerge from their dust cocoon and become T~Tauri stars (TTSs). These pre-main sequence (PMS) stars are still undergoing a phase of contraction towards the main sequence (MS). One distinguishes two kinds of TTSs, the classical T~Tauri stars (cTTSs) accreting material from the circumstellar disc in which planets are in the process of formation, and the weak-line T~Tauri stars (wTTSs), the accretion disc of which has mostly dissipated. TTSs are key targets to constrain theoretical models of stellar and planetary formation. 

In particular, studying the magnetic field of these stars is of crucial importance given the essential role it plays at early stages of stellar evolution, controling the accretion/ejection process at work in cTTSs and therefore angular momentum evolution of TTSs \citep{bouvier07a,frank14}. Thanks to high-resolution spectropolarimeters such as ESPaDOnS \citep{donati03b}, on the 3.6m Canada-France-Hawaii Telescope (CFHT), a small sample of cTTSs has been studied over the last two decades. These studies reveal that cTTSs host strong large-scale magnetic fields of a few hundred gauss to a few kilogauss (e.g. \citealt{donati07,donati11,donati13,alencar12,Bouvier20}), the topology of which mainly reflects the internal structure of the star \citep{donati09,morin10,gregory12}. More specifically, the topology is more complex and departs from a low-order, mainly axisymmetric poloidal field when the star becomes largely radiative.

WTTSs are also targets of interest for further constraining stellar and planetary evolution models, as TTSs evolve from cTTSs to wTTSs and post TTSs. A Large Programme named `Magnetic Topologies of Young Stars and the Survival of close-in giant Exoplanets' (MaTYSSE) carried out with ESPaDOnS, was dedicated to the observation of a few tens of wTTSs to study their large-scale magnetic fields and investigate how different they are compared to those hosted by cTTSs. Studies carried out as part of the MaTYSSE programme showed that most wTTSs follow magnetic trends similar to those of cTTSs, i.e. with strong, simple and mostly axisymmetric large-scale poloidal fields when the star is fully convective. However, some depart from this picture like, e.g., the fully convective stars V410~Tau \citep{yu19,finociety21} and LkCa~4 \citep{donati14} that also exhibit a strong toroidal field.
Another goal of MaTYSSE was to search for close-in massive planets (called hot Jupiters/hJs) orbiting wTTSs, in order to constrain evolutionary models at early stages of planet formation. This effort led to the detection of hJs around 2 wTTSs, namely TAP~26 \citep{yu17} and V830~Tau \citep{donati17}, through the periodic modulation of the radial velocity (RV) of the host star induced by the presence of these massive planets.

New opportunities to investigate the properties of TTSs came with the recently installed near-infrared (NIR) high-resolution spectropolarimeter and high-precision velocimeter SPIRou \citep{donati20} at CFHT. In particular, the SPIRou Legacy Survey (SLS) Large Programme, allocated 310 nights at CFHT, includes a work package dedicated to the study of large-scale magnetic fields of cTTSs and wTTSs, and to the detection of hJs around such stars. As the Zeeman effect is enhanced at NIR wavelengths with respect to the optical domain, SPIRou is especially well suited for magnetic studies of TTSs, that are also often brighter in this spectral range. Moreover, we expect the RV activity jitter induced by the star itself to be smaller in the NIR thanks to the lower brightness contrast between surface features (spots, plages) and the quiet photosphere (e.g. \citealt{mahmud11,crockett12}), making it easier to detect the potential presence of close-in massive planets around these very active stars.

In this paper, we focus on the wTTS LkCa~4 we observed in the framework of SLS, in order to perform an analysis similar to that achieved for V410~Tau \citep{finociety21}. LkCa~4 is a young single fully-convective K7 wTTS \citep{herbig86,grankin13} with a logarithmic luminosity relative to the Sun of $\log{L_*/L_\odot}=-0.04\pm0.11$, an effective temperature and a logarithmic gravity of $T_{\rm eff}=4100\pm50$~K and $\log g = 3.8\pm0.1$ \citep{donati14}, located in the Taurus star-forming region \citep{herbig86,white01,kraus11,grankin13} at a distance of $129.8\pm0.3$~pc \citep{edr3}.This star belongs to the C2-L1495 cloud, whose distance ($129.53$~pc) and age ($1.34\pm0.19$~Myr) were recently estimated from GAIA data \citep{krolikowksi21}. 
Using PMS evolution models of \cite{siess2000}, one finds that LkCa~4 is aged $\unsim2$~Myr, with a mass and a radius of $0.79\pm0.05$~\msun\ and $2.0\pm0.2$~\rsun, respectively \citep{donati14}. Using the more recent evolutionary models of \cite{baraffe15}, we derive for LkCa~4 an age of $\unsim1.3$~Myr, consistent with that of the C2-L1495 cloud, along with a mass of $0.73\pm0.05$~\msun\ and a radius of $1.9\pm0.2$~\rsun. Given the distance, temperature and radius, and the visual extinction derived for this star ($A_V=0.68\pm0.15$, \citealt{donati14}), we infer an unspotted $V$ magnitude of 11.97 for LkCa~4, i.e., significantly brighter than the average observed $V$ magnitude (12.624, \citealt{grankin08}), indicating that the star is heavily spotted.  
A recent study based on optical and NIR spectra further confirms this conclusion, suggesting that as much as 80\% of surface of LkCa~4 may host dark brightness features \citep{gullysantiago17}. 

Based on optical data collected with ESPaDOnS in 2014 in the framework of the MaTYSSE programme, our first study of LkCa~4 \citep{donati14} concluded that the large-scale magnetic field of this star consists of a strong and simple, mostly axisymmetric, poloidal field (featuring a dipole component of 1.6~kG), and of a 1~kG toroidal component. In addition, the reconstructed brightness image features a dark polar spot as well as a warm plage at intermediate latitude that cover altogether about 25\% of the visible surface. These features induce large photometric and velocimetric variations that were reasonably well modeled using tomographic techniques like Zeeman-Doppler Imaging (ZDI; \citealt{semel89,brown91,donatibrown97,donati06}). The results of this study also illustrated that such techniques can mitigate the activity jitter in RV curves up to a RV precision of 0.055~\kms, and suggest that there is no hJ more massive than 1~M$_{\rm jup}$ orbiting LkCa~4 at a distance of 0.1~au or closer (i.e. no planet-induced RV signal with a semi-amplitude larger than 0.1~\kms\ was observed). Finally, \cite{donati14} also reported that the differential rotation (DR) at the surface of LkCa~4 is much lower than that of the Sun, and potentially equal to 0 (solid-body rotation).

As a follow-up to this optical analysis, LkCa~4 has been monitored with SPIRou in the framework of the SLS from 2021 October 14 to 2022 January 30, contemporaneously with the Transiting Exoplanet Survey Satellite (TESS) from 2021 September 16 to November 06, during Sectors 43 and 44. Ground-based photometric measurements were also collected at the Crimean Astrophysical Observatory (CrAO) during the same observing period as SPIRou. We start this paper with a detailed description of our data set (Sec.~\ref{sec:observations}). We then present the results obtained when applying ZDI to recover the brightness distribution and large-scale magnetic field at the surface of the star (Sec.~\ref{sec:tomography}). We investigate the activity of LkCa~4 with velocimetric measurements in Sec.~\ref{sec:radial_velocities} and through the study of three specific lines, known to be activity proxies in the NIR (the \ion{He}{i} triplet at 1083.3 nm, the
Paschen $\beta$ and Brackett $\gamma$ lines) in Sec.~\ref{sec:activity}. We finally summarize and discuss our results in Sec.~\ref{sec:discussion}.

\section{Observations}
\label{sec:observations} % used for referring to this section from elsewhere

\subsection{SPIRou observations}

We observed LkCa~4 with SPIRou, collecting high-resolution spectra ranging from 950 to 2500~nm at a spectral resolving power of $\unsim 70,000$ \citep{donati20} between 2021 Oct 14 and 2022 Jan 30. Our dataset consists of 41 spectropolarimetric observations, each composed of a sequence of 4 subexposures of 550~s taken at different azimuths of the polarimeter retarder in order to remove (to first order) potential sources of spurious polarisation signals and systematic errors \citep{donati97}. These data were reduced using the Libre-ESpRIT pipeline, initally developped for ESPADOnS \citep{donati97}, and adapted for SPIRou observations \citep{donati20}. Telluric correction was performed with a PCA approach similar to that outlined in \cite{artigau14}. This yielded telluric-corrected spectra in both unpolarized (Stokes~$I$) and circularly polarized (Stokes~$V$) spectra, with a signal-to-noise ratio (SNR) per pixel in the $H$ band ranging from 130 to 213 (median of 192). A full journal of observations is given in Table~\ref{tab:journal_observations}.

We applied Least-Square Deconvolution (LSD; \citealt{donati97}) to all our spectra, using a mask generated with the VALD-3 database \citep{vald} and containing only (moderate to strong) atomic lines with known Landé factor, and a relative depth (with respect to the continuum) of at least 3 per cent. This process provided Stokes~$I$ LSD profiles with average noise levels, expressed in units of the unpolarized continuum, ranging from $5.6 \times 10^{-4}$ to $1.7\times 10^{-3}$ (median value of $6.7\times 10^{-4}$) and Stokes~$V$ LSD profiles with average noise levels ranging from $1.8$ to $3.4\times 10^{-4}$ (median of $2.1\times 10^{-4}$). Strong Zeeman signatures are observed in Stokes~$V$ LSD profiles, with peak-to-peak amplitudes of typically 0.5\%. The very obvious distortions in the shape of the Stokes~$I$ profiles indicate that large features are present at the stellar surface, which is further confirmed by the large amplitude of the photometric light curves (see Sec.~\ref{sec:photometric_observations}).

We note that molecular lines located between 1500 and 1800~nm are affected by solar contamination. Applying LSD with a mask containing both atomic and molecular lines therefore does not improve and even degrades the SNRs of the Stokes~$I$ profiles; hence, we did not use such a mask for studying the brightness of LkCa~4 in contrast with what had been done for V410~Tau \citep{finociety21}.

\addtolength{\tabcolsep}{6pt}  
\begin{table*}
\caption{Spectropolarimetric observations of LkCa~4 collected with SPIRou between 2021 October and 2022 January. Columns 1 to 4 list the date, the Coordinated Universal Time, the Barycentric Julian Date and the rotation cycle (computed as indicated in Sec.~\ref{sec:tess}). Columns 5 to 7 give the SNRs of the spectra in the $H$ band, in the Stokes~$I$ and $V$ LSD profiles while column 8 details the equivalent width of the Stokes~$I$ LSD profiles. From column 9 to 13, we list the measured RV, the longitudinal magnetic field, and the activity indicators, named EWVs (see Sec.~\ref{sec:activity}), computed from the \ion{He}{i} triplet at 1083.3~nm, Paschen $\beta$ and Brackett $\gamma$ lines, along with their error bars estimated from photon noise only (the number in parenthesis corresponds to the error bar taking into account intrinsic variability, see Sec.~\ref{sec:activity}).}
\label{tab:journal_observations}
\centering 
\resizebox{\textwidth}{!}{
\begin{threeparttable}
\begin{tabular}{lcccccccccccc}
\\
\hline \hline
Date & UTC & BJD & Cycle & SNR & SNR$_I$ & SNR$_V$ & EW & RV & B$_l$ & \multicolumn{3}{c}{Activity proxies}\\[1mm]
 &  &  &  & &   &  &  &  &  & \ion{He}{i} & Pa$\beta$ & Br$\gamma$ \\
 
  &  & 2459000+ &  & &  &  & (\kms)  & (\kms) & (G) & (pm) & (pm) & (pm) \\ \hline
 
 2021 October 14 & 12:34:10 & 502.024 & 0.000 & 185 & 1285 & 4405 & 1.166 ± 0.016 & 2.000 ± 0.176 & 44 ± 16 & 3.9 ± 0.7 (5.4) & 3.2 ± 0.8 (1.9) & -7.5 ± 2.6 (6.6) \\ 
2021 October 15 & 12:11:33 & 503.008 & 0.292 & 162 & 1094 & 3975 & 1.015 ± 0.016 & -0.669 ± 0.209 & 227 ± 18 & -2.2 ± 0.7 (5.4) & -3.6 ± 0.8 (1.9) & -5.1 ± 2.6 (6.6) \\ 
2021 October 19 & 13:49:55 & 507.076 & 1.498 & 205 & 1522 & 5180 & 1.341 ± 0.015 & -1.327 ± 0.143 & 135 ± 13 & -11.6 ± 0.7 (5.4) & -3.7 ± 0.8 (1.9) & -5.6 ± 2.6 (6.6) \\ 
2021 October 20 & 12:48:26 & 508.034 & 1.782 & 191 & 1503 & 4851 & 1.415 ± 0.015 & 0.766 ± 0.135 & 227 ± 13 & -9.9 ± 0.7 (5.4) & 1.2 ± 0.8 (1.9) & -4.9 ± 2.6 (6.6) \\ 
2021 October 21 & 12:52:10 & 509.036 & 2.080 & 173 & 1365 & 3898 & 1.170 ± 0.015 & -0.097 ± 0.172 & 182 ± 19 & 3.5 ± 0.7 (5.4) & 2.5 ± 0.8 (1.9) & -6.5 ± 2.6 (6.6) \\ 
2021 October 22 & 12:52:41 & 510.037 & 2.376 & 145 & 1282 & 3631 & 1.289 ± 0.017 & 0.047 ± 0.172 & 184 ± 19 & -18.9 ± 0.7 (5.4) & -3.7 ± 0.8 (1.9) & -4.7 ± 2.6 (6.6) \\ 
2021 October 23 & 12:35:50 & 511.025 & 2.669 & 199 & 1614 & 5295 & 1.456 ± 0.014 & -1.163 ± 0.131 & 226 ± 12 & -8.8 ± 0.7 (5.4) & -1.0 ± 0.8 (1.9) & -2.9 ± 2.6 (6.6) \\ 
2021 October 24 & 13:17:42 & 512.054 & 2.975 & 154 & 1387 & 3813 & 1.256 ± 0.016 & 2.621 ± 0.165 & 47 ± 19 & 5.0 ± 0.7 (5.4) & 6.0 ± 0.8 (1.9) & -2.9 ± 2.6 (6.6) \\ 
2021 October 25 & 08:44:44 & 512.864 & 3.215 & 194 & 1486 & 4517 & 1.255 ± 0.015 & -1.313 ± 0.150 & 252 ± 16 & 1.1 ± 0.7 (5.4) & -2.9 ± 0.8 (1.9) & -6.3 ± 2.6 (6.6) \\ 
2021 October 26 & 11:57:48 & 513.998 & 3.551 & 199 & 1575 & 5157 & 1.402 ± 0.015 & -1.607 ± 0.140 & 152 ± 13 & -10.4 ± 0.7 (5.4) & -3.6 ± 0.8 (1.9) & -2.1 ± 2.6 (6.6) \\ 
2021 October 28 & 13:15:21 & 516.052 & 4.160 & 169 & 1455 & 4323 & 1.262 ± 0.016 & -1.779 ± 0.164 & 228 ± 17 & -2.0 ± 0.7 (5.4) & 1.5 ± 0.8 (1.9) & -2.5 ± 2.6 (6.6) \\ 
2021 November 18 & 13:51:48 & 537.078 & 10.396 & 188 & 1618 & 4597 & 1.378 ± 0.014 & -0.264 ± 0.133 & 176 ± 15 & -7.9 ± 0.7 (5.4) & -1.1 ± 0.8 (1.9) & -0.0 ± 2.6 (6.6) \\ 
2021 November 19 & 12:16:07 & 538.011 & 10.672 & 210 & 1563 & 5203 & 1.393 ± 0.015 & -2.637 ± 0.132 & 217 ± 12 & -1.1 ± 0.7 (5.4) & -1.7 ± 0.8 (1.9) & -3.1 ± 2.6 (6.6) \\ 
2021 November 20 & 10:58:17 & 538.957 & 10.953 & 176 & 1611 & 4367 & 1.334 ± 0.014 & 2.960 ± 0.142 & 47 ± 16 & 4.8 ± 0.7 (5.4) & 5.9 ± 0.8 (1.9) & -2.1 ± 2.6 (6.6) \\ 
2021 November 21 & 11:06:06 & 539.963 & 11.251 & 196 & 1590 & 4939 & 1.330 ± 0.014 & -1.245 ± 0.140 & 230 ± 14 & -2.5 ± 0.7 (5.4) & -3.4 ± 0.8 (1.9) & -3.4 ± 2.6 (6.6) \\ 
2021 November 22 & 11:39:43 & 540.986 & 11.555 & 200 & 1673 & 5173 & 1.467 ± 0.014 & -1.392 ± 0.124 & 160 ± 13 & -8.9 ± 0.7 (5.4) & -2.7 ± 0.8 (1.9) & -5.3 ± 2.6 (6.6) \\ 
2021 December 9 & 10:52:45 & 557.953 & 16.586 & 192 & 1618 & 4161 & 1.462 ± 0.014 & -1.676 ± 0.130 & 168 ± 16 & -3.0 ± 0.7 (5.4) & -0.2 ± 0.8 (1.9) & 2.6 ± 2.6 (6.6) \\ 
2021 December 10 & 10:15:39 & 558.928 & 16.875 & 191 & 1603 & 4788 & 1.385 ± 0.014 & 2.357 ± 0.133 & 156 ± 14 & 4.1 ± 0.7 (5.4) & -0.2 ± 0.8 (1.9) & 0.9 ± 2.6 (6.6) \\ 
2021 December 11 & 10:21:37 & 559.932 & 17.173 & 138 & 1441 & 3301 & 1.270 ± 0.015 & -1.817 ± 0.159 & 217 ± 22 & 1.0 ± 0.7 (5.4) & -3.1 ± 0.8 (1.9) & -2.8 ± 2.6 (6.6) \\ 
2021 December 12 & 10:29:22 & 560.937 & 17.471 & 143 & 1613 & 3552 & 1.410 ± 0.014 & -0.680 ± 0.133 & 104 ± 19 & -1.5 ± 0.7 (5.4) & -1.7 ± 0.8 (1.9) & -1.4 ± 2.6 (6.6) \\ 
2021 December 14 & 13:25:28 & 563.059 & 18.101 & 158 & 1466 & 3620 & 1.281 ± 0.015 & -0.254 ± 0.164 & 191 ± 20 & 2.7 ± 0.7 (5.4) & -1.6 ± 0.8 (1.9) & 5.3 ± 2.6 (6.6) \\ 
2021 December 15 & 08:07:13 & 563.838 & 18.332 & 152 & 1497 & 3573 & 1.374 ± 0.015 & -0.172 ± 0.145 & 223 ± 19 & -11.9 ± 0.7 (5.4) & -6.3 ± 0.8 (1.9) & 1.4 ± 2.6 (6.6) \\ 
2021 December 16 & 09:45:41 & 564.907 & 18.649 & 169 & 1623 & 4215 & 1.483 ± 0.014 & -1.584 ± 0.128 & 200 ± 15 & -3.0 ± 0.7 (5.4) & -0.8 ± 0.8 (1.9) & 0.4 ± 2.6 (6.6) \\ 
2021 December 18 & 10:11:50 & 566.925 & 19.247 & 157 & 1114 & 3661 & 1.208 ± 0.018 & -1.250 ± 0.202 & 231 ± 19 & -3.1 ± 0.7 (5.4) & -6.7 ± 0.8 (1.9) & -0.3 ± 2.6 (6.6) \\ 
2022 January 6 & 10:01:22 & 585.918 & 24.880 & 213 & 1139 & 5683 & 1.463 ± 0.021 & 2.869 ± 0.191 & 136 ± 12 & 6.7 ± 0.7 (5.4) & 3.8 ± 0.8 (1.9) & -2.3 ± 2.6 (6.6) \\ 
2022 January 8 & 09:41:27 & 587.904 & 25.469 & 197 & 1516 & 5119 & 1.386 ± 0.015 & 0.121 ± 0.142 & 144 ± 13 & -2.4 ± 0.7 (5.4) & 0.8 ± 0.8 (1.9) & 3.8 ± 2.6 (6.6) \\ 
2022 January 9 & 09:43:21 & 588.905 & 25.766 & 203 & 1773 & 5150 & 1.478 ± 0.013 & 0.189 ± 0.111 & 236 ± 12 & 3.5 ± 0.7 (5.4) & 0.6 ± 0.8 (1.9) & 2.9 ± 2.6 (6.6) \\ 
2022 January 10 & 10:53:03 & 589.954 & 26.076 & 195 & 1630 & 4628 & 1.274 ± 0.014 & 0.133 ± 0.145 & 162 ± 16 & -1.0 ± 0.7 (5.4) & 5.6 ± 0.8 (1.9) & 2.2 ± 2.6 (6.6) \\ 
2022 January 11 & 06:00:12 & 590.750 & 26.313 & 209 & 1740 & 5399 & 1.423 ± 0.013 & -0.220 ± 0.130 & 211 ± 13 & -3.8 ± 0.7 (5.4) & -3.7 ± 0.8 (1.9) & -4.3 ± 2.6 (6.6) \\ 
2022 January 12 & 10:19:13 & 591.930 & 26.663 & 205 & 1588 & 5029 & 1.474 ± 0.015 & -1.348 ± 0.128 & 206 ± 13 & -1.3 ± 0.7 (5.4) & -0.2 ± 0.8 (1.9) & 3.0 ± 2.6 (6.6) \\ 
2022 January 13 & 09:46:35 & 592.907 & 26.952 & 195 & 919 & 5014 & 0.846 ± 0.016 & 3.462 ± 0.245 & 86 ± 14 & -0.3 ± 0.7 (5.4) & 3.4 ± 0.8 (1.9) & 13.7 ± 2.6 (6.6) \\ 
2022 January 15 & 10:15:20 & 594.927 & 27.551 & 207 & 1502 & 5161 & 1.419 ± 0.015 & -0.718 ± 0.151 & 137 ± 13 & 2.8 ± 0.7 (5.4) & 1.2 ± 0.8 (1.9) & 4.9 ± 2.6 (6.6) \\ 
2022 January 18 & 09:43:56 & 597.906 & 28.435 & 205 & 1100 & 4909 & 0.999 ± 0.015 & 0.509 ± 0.195 & 152 ± 14 & 4.1 ± 0.7 (5.4) & -0.4 ± 0.8 (1.9) & 12.1 ± 2.6 (6.6) \\ 
2022 January 19 & 09:08:20 & 598.881 & 28.724 & 208 & 1282 & 5238 & 1.267 ± 0.015 & -0.503 ± 0.156 & 254 ± 12 & 3.0 ± 0.7 (5.4) & -1.2 ± 0.8 (1.9) & 6.1 ± 2.6 (6.6) \\ 
2022 January 20 & 08:50:05 & 599.868 & 29.017 & 199 & 641 & 5342 & 1.209 ± 0.037 & 0.740 ± 0.368 & 75 ± 14 & -0.7 ± 0.7 (5.4) & 5.1 ± 0.8 (1.9) & 12.3 ± 2.6 (6.6) \\ 
2022 January 21 & 08:56:25 & 600.873 & 29.315 & 205 & 1041 & 4678 & 0.952 ± 0.015 & 0.265 ± 0.225 & 182 ± 15 & 0.4 ± 0.7 (5.4) & -2.6 ± 0.8 (1.9) & 8.6 ± 2.6 (6.6) \\ 
2022 January 22 & 08:02:07 & 601.835 & 29.600 & 130 & 1235 & 2938 & 1.451 ± 0.019 & -1.185 ± 0.159 & 177 ± 22 & 1.1 ± 0.7 (5.4) & 2.4 ± 0.8 (1.9) & 3.2 ± 2.6 (6.6) \\ 
2022 January 24 & 09:09:48 & 603.882 & 30.207 & 196 & 682 & 5684 & 1.306 ± 0.036 & -1.913 ± 0.336 & 200 ± 13 & 15.8 ± 0.7 (5.4) & 0.5 ± 0.8 (1.9) & 9.0 ± 2.6 (6.6) \\ 
2022 January 25 & 09:17:16 & 604.887 & 30.505 & 169 & 1032 & 4177 & 1.462 ± 0.023 & -0.432 ± 0.212 & 117 ± 16 & 2.1 ± 0.7 (5.4) & 4.3 ± 0.8 (1.9) & 5.4 ± 2.6 (6.6) \\ 
2022 January 28 & 08:31:31 & 607.855 & 31.385 & 172 & 1220 & 3723 & 1.509 ± 0.020 & -0.355 ± 0.183 & 200 ± 18 & -1.6 ± 0.7 (5.4) & -0.1 ± 0.8 (1.9) & 0.7 ± 2.6 (6.6) \\ 
2022 January 30 & 08:28:11 & 609.853 & 31.978 & 182 & 576 & 3962 & 1.116 ± 0.038 & 0.963 ± 0.418 & 102 ± 18 & 3.9 ± 0.7 (5.4) & 3.7 ± 0.8 (1.9) & 27.1 ± 2.6 (6.6) \\    \hline

\end{tabular}

%\begin{tablenotes}
%\item $^{(a)}$
%\end{tablenotes}

\end{threeparttable}
}
\end{table*}
\addtolength{\tabcolsep}{-6pt}

\subsection{Photometric observations}
\label{sec:photometric_observations}

\subsubsection{TESS light curve}
\label{sec:tess}

LkCa~4 (TIC 58108662) was also observed by TESS \citep{ricker14}, with a cadence of 2~min, during the monitoring of Sectors 43 (2021 Sep 16 -- Oct 12) and 44 (2021 Oct 12 -- Nov 06) over a total time span of 49~d. The second monitoring was contemporaneous with SPIRou observations, although both instruments are sensitive to different wavelengths (the TESS filter being centred close to the mean wavelength of the $I_{\rm c}$ band).

The TESS data were processed by the Science Processing Operations Center (SPOC; \citealt{jenkins16}) data pipeline (version 5.0). For each sector, the observations were stopped for about 1~d between two physical orbits of the telescope to download the data, yielding a total of about 46~d of science observations\footnote{Details on the processing of the data can be found in the Data Release notes of Sector 43 and 44 (DR62 and DR64) available at \url{https://archive.stsci.edu/tess/tess_drn.html}.}. As we want to use these data to characterise stellar variability, we kept the Pre-search Data Conditioning Single Aperture Photometry (PDCSAP) flux, already corrected for instrumental variations and contamination from nearby stars \citep{smith12,stumpe12,stumpe14}, that was not flagged by the SPOC pipeline. 

\begin{figure}
    \centering
    \includegraphics[scale=0.21,trim={1cm 0.5cm 1cm 1cm},clip]{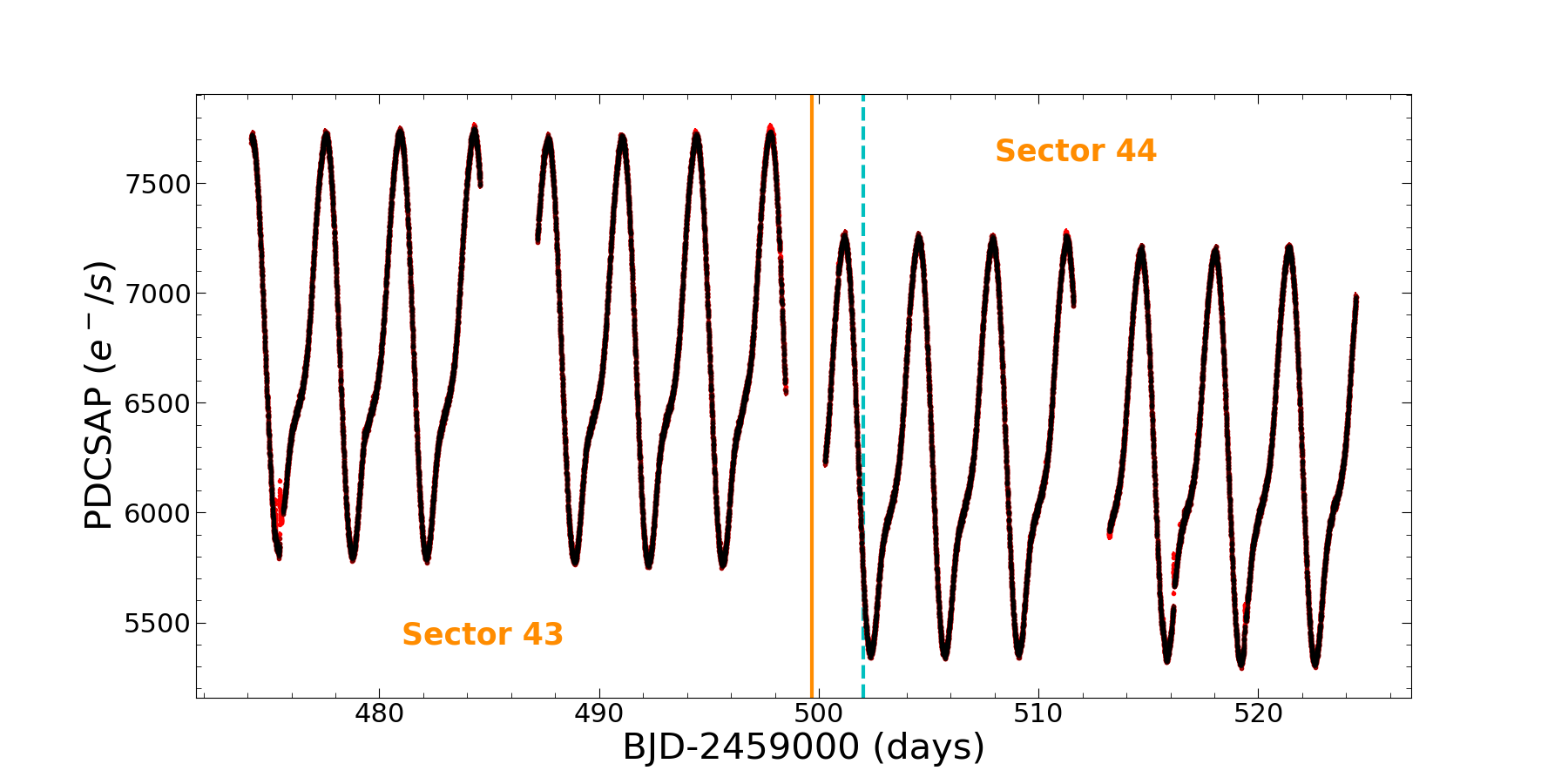}
    \caption{TESS Pre-search Data Conditioning Single Aperture Photometry. The black dots correspond to the filtered light curve while the red ones are those that were either rejected by our $3\sigma$-clipping process or flagged by the SPOC pipeline. The cyan dashed vertical line denotes the start of the SPIRou monitoring while the orange vertical line delimits the light curve of Sector 43 from that of Sector 44.}
    \label{fig:TESS_light_curve}
\end{figure}

We filtered the light curves from flares by applying the same 3$\sigma$-clipping process involving Gaussian Process Regression (GPR; \citealt{rasmussen06}) as in \cite{finociety21}. The obtained filtered light curve (see Fig.~\ref{fig:TESS_light_curve}) was then modelled using a Gaussian Process (GP) with a quasi-periodic kernel (Eq.~\eqref{eq:QP_kernel}) known to be adapted for the description of signals induced by stellar activity \citep{rajpaul15} :

\begin{equation}
    k(t,t') = \theta_1^2 \, \exp\left[-\frac{(t-t')^2}{2\,\theta_2^2} - \frac{\sin^2\left( \frac{(t-t')\pi}{\theta_3}\right)}{2\,\theta_4^2}\right]
    \label{eq:QP_kernel}
\end{equation}

where $t$ and $t'$ are the dates associated with two different observations. $\theta_1$ corresponds to the amplitude of the GP, $\theta_2$ represents the exponential decay time-scale (giving an estimate of the typical spot lifetime), $\theta_3$ is the recurrence time-scale (expected to be equal, or very close, to the stellar rotation period) and $\theta_4$ is the smoothing parameter controlling the short-term variations that are included in the fit.

We find that the decay time-scale is equal to $150\pm37$~d, indicating that the light curve is very close to a purely periodic signal over the time interval covered by the TESS data (i.e. the spot distribution does not evolve much over the time span of the TESS observations). We also estimate $\theta_3=3.372\pm0.002$~d, thus consistent with the stellar rotation period of 3.374~d previously reported by \cite{grankin08} and $\theta_4=0.612 \pm 0.096$. We see a vertical jump in the light curve between Sectors 43 and 44 ($\rm BJD \approx 2459500$), likely due to instrumental effects. As part of the process, we therefore derived the best constant mean value of each Sector to model the flux with the GP, found to be equal to $m_1=6584 \pm 417$~$\mathrm{e^-/s}$ and $m_2=6154\pm417$~$\mathrm{e^-/s}$ for Sector 43 and 44, respectively. We also notice a vertical shift between the first and second half of the TESS light curve associated with Sector 44 (around $\rm BJD = 2459515$), that we corrected (by increasing the lower section) to ensure that both parts of the curve have the same mean level.

We used the stellar rotation period derived from our filtered TESS light curve to compute the rotational cycles $c$ according to:

\begin{equation}
    \mathrm{BJD\;(d)} = 2459502.02 + 3.372\,c
\end{equation}

where the initial date ($\rm BJD_0 =2459502.02$) was arbitrarily chosen to correspond to the date of our first SPIRou observation of LkCa~4.

The TESS monitoring of Sector 44 being contemporaneous with SPIRou observations, we also computed relative photometry (with respect to the mean value $m_2$ derived through GPR) and median time every 20 points, resulting in 741 photometric data points to be included in the imaging process, for which we set the uncertainty to 2.4~mmag (see Sec.~\ref{sec:tomography}), i.e. equal to the RMS dispersion of the selected TESS data about the GPR fit, and slightly larger than the nominal error bar of the TESS data, of 1.9~mmag. 

\subsubsection{Ground-based observations}
\label{sec:ground_based_observations}

Fourteen additional photometric observations were collected in the $V$, $R_{\rm j}$, $R_{\rm c}$, $I_{\rm j}$ and $I_{\rm c}$ bands with the ground-based CrAO 1.25-m AZT-11 telescope from 2021 Oct 10 to 2022 Jan 27. The full log of these observations can be found in Table~\ref{tab:log_crao}.

\begin{figure}
    \centering
    \includegraphics[scale=0.32,trim={0.5cm 1.8cm 3cm 3cm},clip]{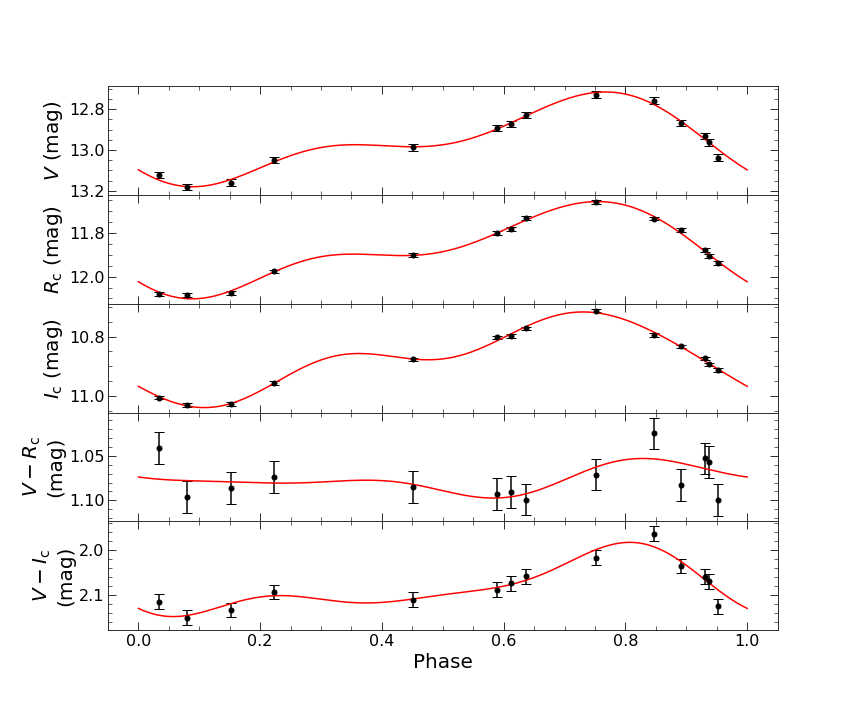}
    \caption{Magnitudes in the $V$, $R_{\rm c}$, $I_{\rm c}$ bands (1$^{\rm st}$ to $3^{\rm rd}$ panels) and $V-R_{\rm c}$ and $V-I_{\rm c}$ colour indexes ($4^{\rm th}$ and $5^{\rm th}$ panels) collected with the ground-based AZT-11 telescope at the CrAO between 2021 October and 2022 January. For the three first panels, the red line corresponds to a fit to the data involving a sine wave and the first two harmonics. The empirical error bars on the magnitudes are then set to ensure a unit $\chi^2_r$ between the data and these models (16, 9 and 6~mmag for the $V$, $R_{\rm c}$ and $I_{\rm c}$ bands, respectively). For the last two panels, the red curves depict the difference between the previous models. All light curves are phased using the ephemeris of Sec.~\ref{sec:tess}. }
    \label{fig:crao_observations}
\end{figure}

These measurements also show a brightness modulation, that can be fitted with a periodic function (sine wave plus two harmonics) as shown in Fig.~\ref{fig:crao_observations}, assuming the stellar rotation period derived from the TESS light curve. As no nominal uncertainties are available, we estimated empirical error bars by setting them to ensure a unit reduced chi-squared ($\chi^2_r$) between the measurements and the model. Using only the magnitudes in the $V$, $R_{\rm c}$ and $I_{\rm c}$ bands, we found typical uncertainties of 16, 9 and 6~mmag, for each band respectively. As expected, we see that the amplitude of the light curve decreases with wavelength and is equal to $0.464\pm0.020$, $0.447\pm0.011$ and $0.323\pm0.008$~mag in the $V$, $R_{\rm c}$ and $I_{\rm c}$ bands, the latter being consistent with the TESS light curve featuring a full amplitude of $0.3207\pm0.0006$~mag. The observed photometric modulation clearly reflects the presence of large surface inhomogeneities coming in and going out of the observer's view as the star rotates.

\begin{figure}
    \centering
    \includegraphics[scale=0.27,trim={0cm .5cm 1cm 1cm},clip]{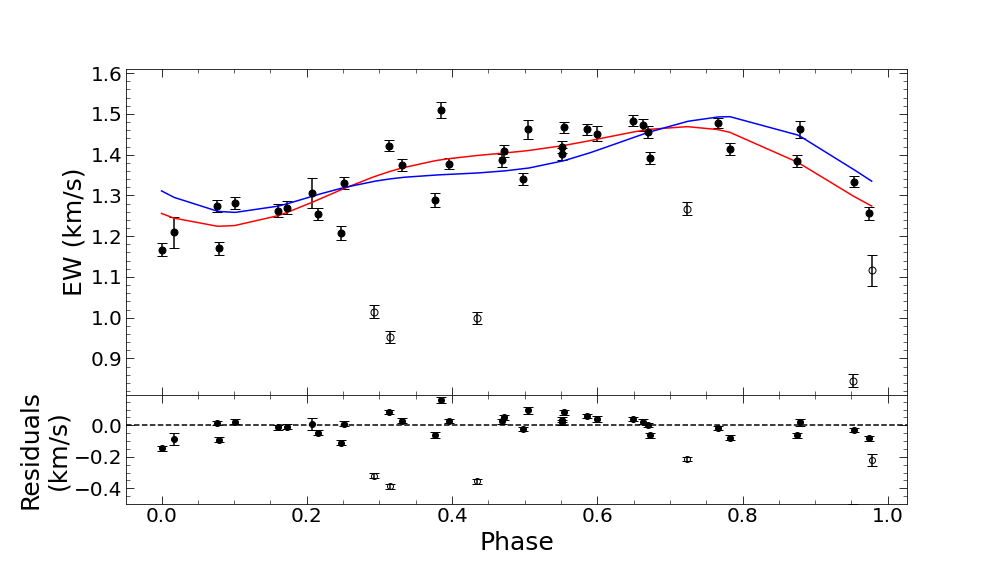}
    \caption{Phase-folded EWs of the Stokes~$I$ LSD profiles. \textit{Top panel:} The measurements along with their error bars are shown as black dots. The ZDI fit is displayed as a blue line while the red curve corresponds to a periodic fit (including the fundamental and the first harmonic) to the measurements associated with a full circle (the open circles being rejected through a $\sigma$-clipping process). Both models show a full amplitude of $\unsim0.25$~\kms, corresponding to a peak-to-peak variation of 20\%. \textit{Bottom panel:} Residuals between the measurements and the ZDI fit, with a RMS dispersion of 0.06~\kms\ (when excluding the open circles).}
    \label{fig:EW_modulation}
\end{figure}

\begin{figure*}
    \centering
    \includegraphics[scale=0.18,trim={0cm 0cm 0cm 0cm},clip]{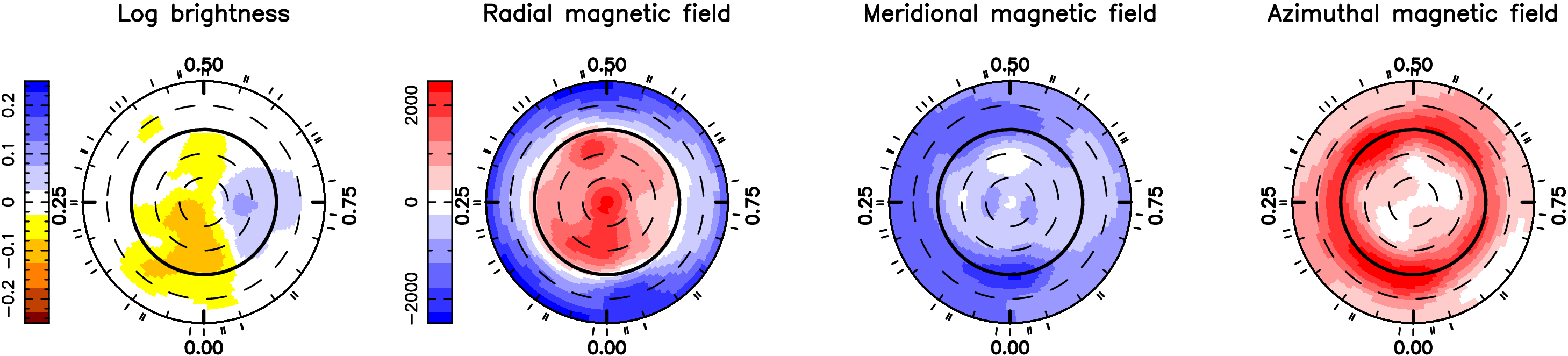}
    \caption{ZDI maps of the logarithmic brightness surface (left panel), radial, meridional and azimuthal magnetic field components (second to fourth panels). The star is shown in a flattened polar view, with the pole at the center, the equator as a bold circle and the 30$^\circ$ and $60^\circ$ latitude paralleles depicted in dashed circles. The star is shown down to -60$^\circ$ below which almost nothing contributes to the data. For the brightness map, yellow/brown corresponds to dark spots while blue corresponds to warm plages. For the magnetic maps, red is associated to a radial, meridional and azimuthal field pointing outwards, polewards and counter-clockwise, respectively. The ticks around the star refer to the phases of the 41 spectropolarimetric observations collected with SPIRou.}
    \label{fig:zdi_maps}
\end{figure*}

\begin{figure*}
    \centering
    \begin{subfigure}{0.49\textwidth}
         \centering
         \includegraphics[scale=0.15,trim={0cm 0cm 0cm 0cm},clip]{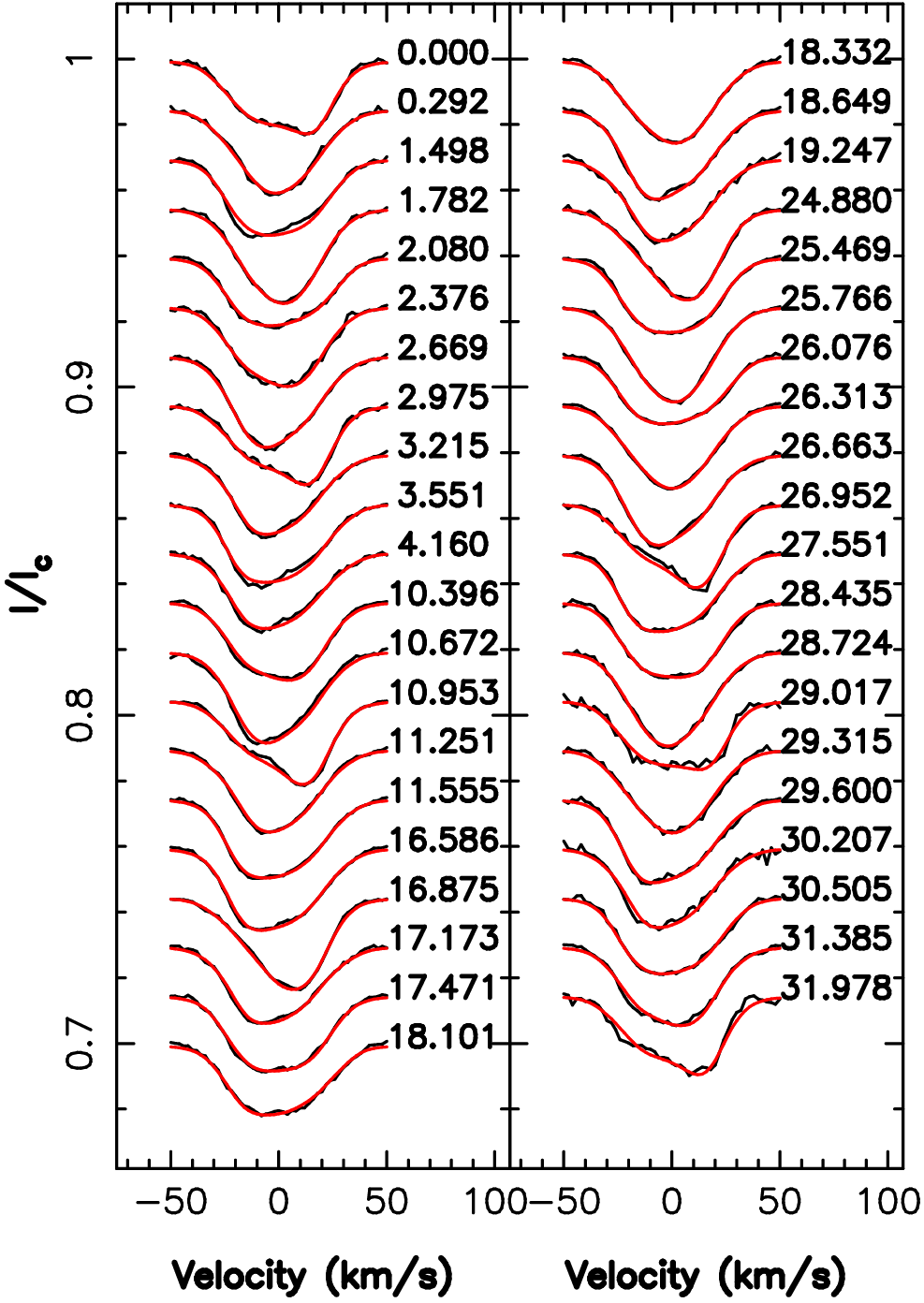}
         
    \end{subfigure}
    \hfill
    %\hspace*{-0.4cm}
    \begin{subfigure}{0.49\textwidth}
         \centering
         \includegraphics[scale=0.15,trim={0cm 0cm 0cm 0cm},clip]{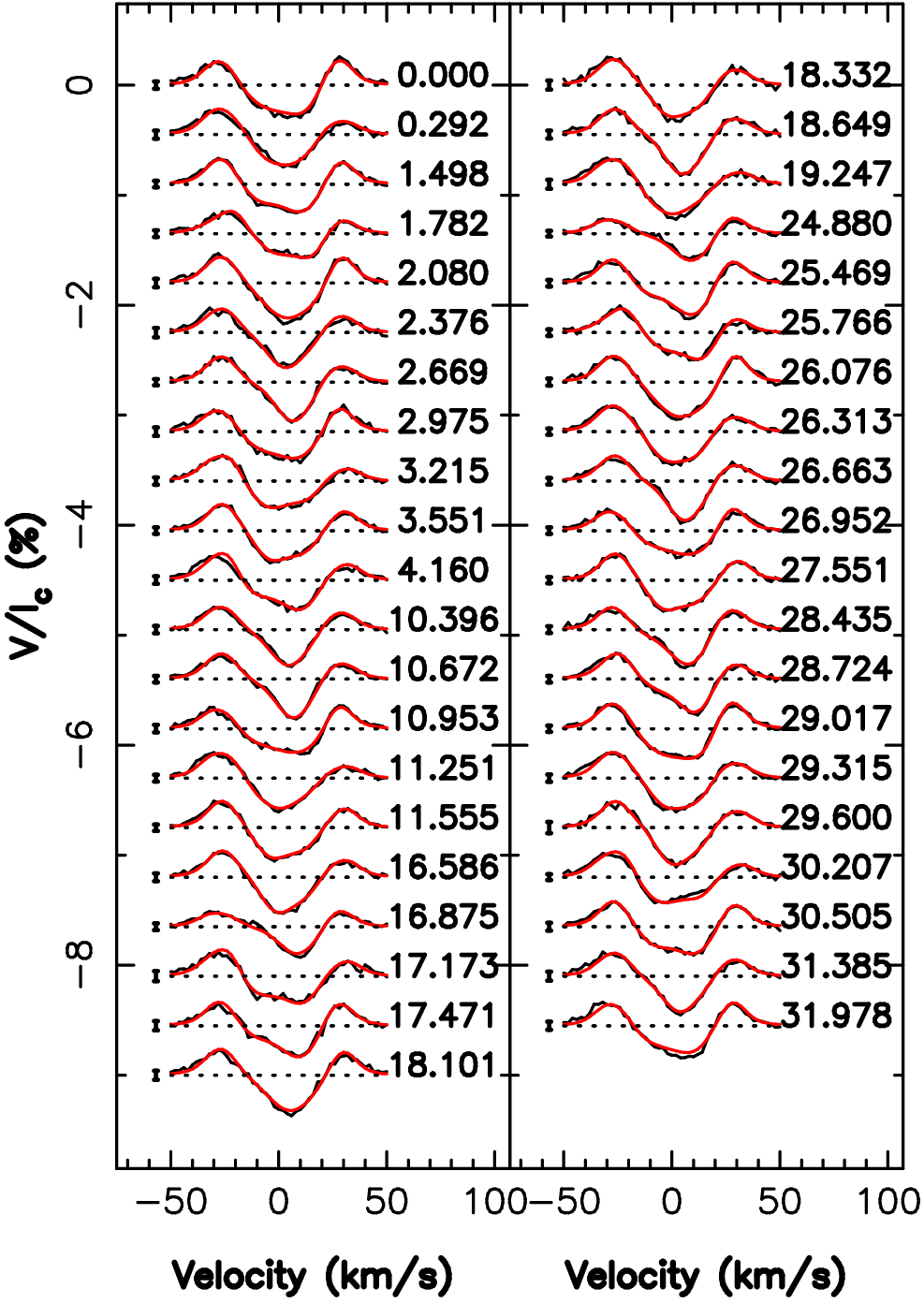}
         
    \end{subfigure}
    \caption{Stokes~$I$ (left) and $V$ (right) LSD profiles. For both panels, the observed Stokes profiles are shown in black while the ZDI fit is depicted as a red line. The rotation cycle associated with each observation is mentioned on the right of each profile. For Stokes~$V$ LSD profiles, we also display the $3\sigma$ error bars on the left of each profile.}
    \label{fig:LSD_profiles}
\end{figure*}

\section{Tomographic imaging}
\label{sec:tomography}
ZDI \citep{semel89,brown91,donatibrown97,donati06,donati14} is an efficient tomographic technique inspired from medical applications allowing one to recover the brightness distribution and large-scale magnetic topology at the surface of an active star. We applied this technique on both our sets of Stokes~$I$ and Stokes~$V$ LSD profiles to recover brightness and magnetic maps of LkCa~4. 

ZDI is an iterative process based on the principles of maximum entropy that iteratively inverts sets of Stokes~$I$ and $V$ LSD profiles, by adding brightness inhomogeneities and magnetic regions at the surface of the star and comparing the associated LSD profiles with the observed ones until reaching a unit $\chi^2_r$. 
In practice, the stellar surface is divided in 3000 cells; we then compute the local Stokes~$I$ and $V$ LSD profiles from each cell using the Unno-Rachkovsky's solution of the polarized radiative transfer equations in a plane-parallel Milne-Eddington atmosphere (e.g. \citealt{landi04}). The built-in prescription for the limb-darkening variations has been replaced by a linear law for the continuum only associated with a coefficient $\epsilon=0.3$, consistent with $T_{\rm eff}=4100$~K and $\log g = 4.0$ \citep{claret11}. The synthetic LSD profiles are then derived by integrating all local profiles over the visible stellar hemisphere. 

The relative brightness is simply described as a set of independent values for each grid cell at the surface of the star; the large-scale magnetic field is expressed as the sum of a poloidal and a toroidal component, both described as spherical harmonic expansions (\citealt{donati06,finociety22}).

As the equivalent width (EW) of our Stokes~$I$ LSD profiles is found to be modulated with rotation phase (by about 20\% peak-to-peak, see Fig.~\ref{fig:EW_modulation}), we added an empirical description of how the local profile varies with temperature. In practice, we simply assume that the depth of the local profile varies as a power $\delta$ of the local brightness (over the limited range of brightness values that we reconstruct, see Sec.~\ref{sec:brightness_magnetic}). Since the EWs of Stokes $I$ profiles are smaller when the star is fainter, it already indicates that $\delta$ is positive.

Given the small amount of surface differential rotation reported for LkCa~4 \citep{donati14}, we will first consider that the star rotates as a solid body before estimating the differential rotation parameters using our data in Sec~\ref{sec:differential_rotation}.

We fitted our LSD profiles using a line model featuring a mean wavelength, Doppler width and Landé factor of 1750~nm, 3.4~\kms and 1.2, respectively. We set the inclination $i=70^\circ$ and line-of-sight projected equatorial velocity $v\sin{i}=28$~\kms\ as in \cite{donati14}. As part of the imaging process, we also retrieve a bulk radial velocity of $16.9\pm0.1$~\kms, consistent with the value reported in \cite{donati14}. Table~\ref{tab:stellar_parameters} gathers all physical parameters of LkCa~4.

\begin{table}
    \caption{Physical parameters of LkCa~4. From top to bottom: distance from Earth, bulk radial velocity, effective temperature, rotation period, luminosity, inclination, line-of-sight projected equatorial velocity, minimal stellar radius, stellar radius, stellar mass, logarithmic surface gravity, rotation rate at the equator, pole-to-equator rotation rate difference and age. }
    \label{tab:stellar_parameters}
    \centering 
    \begin{tabular}{lll} \hline
         Parameter & Value & Reference  \\\hline
         d (pc) & $129.8\pm0.3$ & \cite{edr3} \\
         Bulk RV (\kms) & $16.9\pm0.8$ & This work (Sec.~\ref{sec:tomography}) \\
         $T_{\rm eff}$ (K) & $4100\pm50$ & \cite{donati14} \\
         $P_{\rm rot}$ (d)  & $3.372\pm0.002$ & This work (Secs.~\ref{sec:tess} \& \ref{sec:radial_velocities} ) \\
         $\log{L_*/L_\odot}$ & $-0.04\pm0.11$ & \cite{donati14} \\
         $i$ ($^\circ$) & 70 & \cite{donati14} \\
         $v\sin{i}$ (\kms) & $28.0\pm0.5$ & \cite{donati14} \\
         R$_* \sin{i}$ (R$_\odot$) & $1.87\pm0.03$ & \cite{donati14} \\
         R$_*$ (R$_\odot$) & $2.0\pm0.2$ & \cite{donati14} \\
         M$_*$ (\msun) & $0.73\pm0.05$ & This work \\
         $\log{g}$ (cgs units) & $3.8\pm0.1$ & \cite{donati14} \\
         $\Omega_{\rm eq}$ ($\rm mrad\,d^{-1}$) & $1864.0\pm0.2$ & This work (Sec.~\ref{sec:differential_rotation}) \\
         d$\Omega$ ($\rm mrad\,d^{-1}$) & $5.8\pm0.8$ & This work (Sec.~\ref{sec:differential_rotation}) \\
         Age (Myr) & $\unsim 1.3$ & This work \\\hline
    \end{tabular}
\end{table}

\subsection{Brightness and magnetic imaging}
\label{sec:brightness_magnetic}

In this Section, we assume that the star rotates as a solid body. We first simultaneously inverted both our Stokes~$I$ and $V$ LSD profiles using ZDI without including the TESS data in the fitting process. Both sets of LSD profiles were adjusted down to $\chi^2_r=1$. We show the reconstructed brightness and magnetic maps in Fig.~\ref{fig:zdi_maps} with the associated profiles in Fig.~\ref{fig:LSD_profiles}. At this stage, adjusting $\delta$, so that the modulation with rotation phase of the EW of the synthetic LSD profiles matched observations, was essential, which yielded $\delta=1.0\pm 0.1$. This ensures at the same time that the amplitude of the predicted light curve is consistent with that observed with TESS and at CrAO (once the reconstructed brightness image in the $H$ band is translated into an $I_{\rm c}$ band image using the Planck function, assuming a photospheric temperature of $4100$~K).

We note that the brightness distribution is mainly characterized by two large structures covering about 7\% of the stellar surface: a dark spot at phase 0.1 spreads between the pole and the equator whereas a bright plage is reconstructed at phase 0.75. These features are the ones that generate most of the observed rotational modulation of both the light curve and the EWs of LSD Stokes $I$ profiles. The reconstructed spot coverage is much lower than that derived by \cite{gullysantiago17}, reaching up to 80\%, which implies that the star is most likely evenly covered with small-scale cool spots that are not resolved by ZDI.

\begin{figure}
    \centering
    %\hspace*{-0.4cm}
    \begin{subfigure}{0.49\textwidth}
         \centering
         \includegraphics[scale=0.11,trim={0cm 0cm 0cm 0cm},clip]{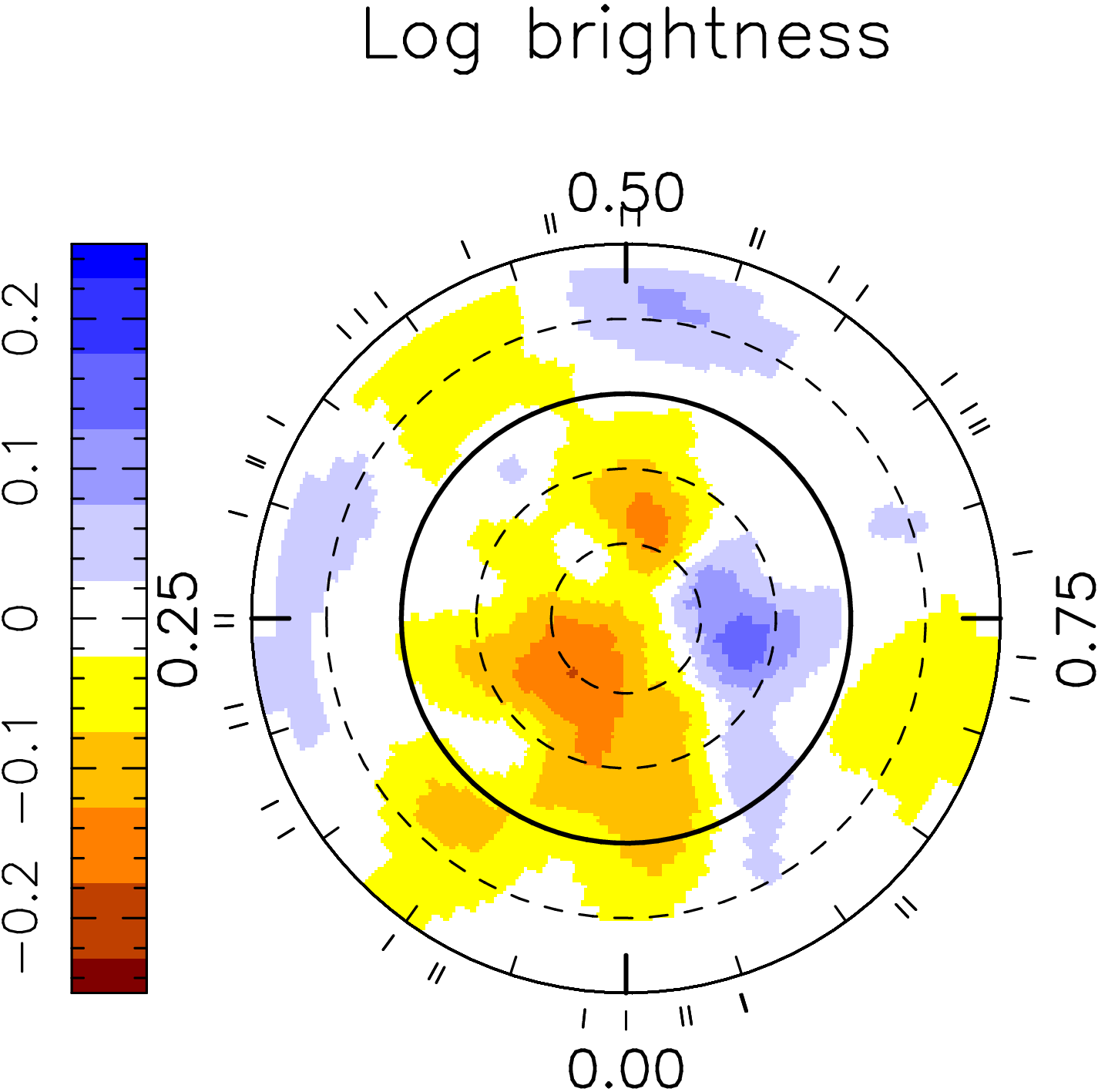}
         
    \end{subfigure}
    \hfill
    \hspace*{-0.4cm}
    \begin{subfigure}{0.49\textwidth}
         \centering
         \includegraphics[scale=0.27,trim={0cm 0cm 3.5cm 2cm},clip]{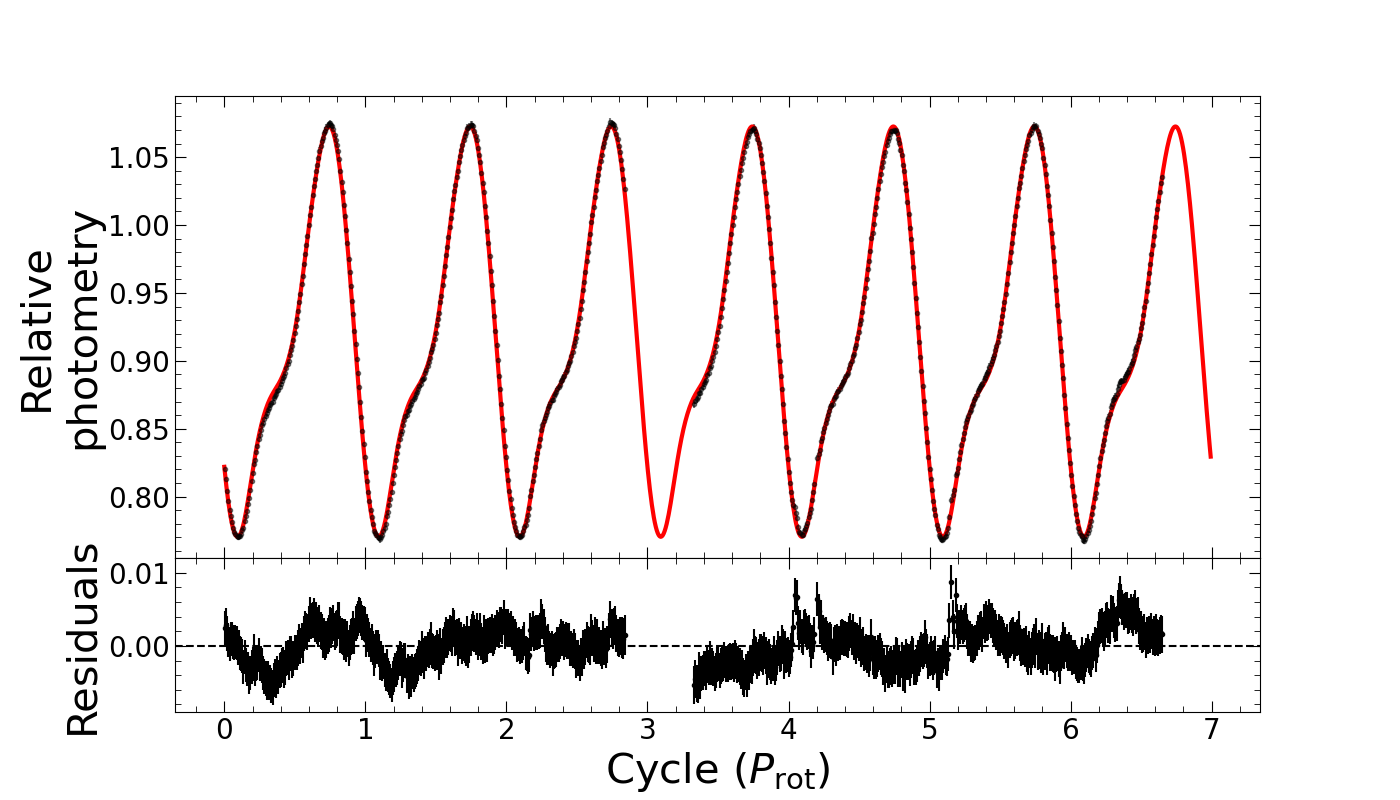}
         
    \end{subfigure}

    \caption{ZDI reconstruction using SPIRou and TESS data simultaneously. \textit{Top panel:} Map of the logarithmic relative brightness as described in Fig.~\ref{fig:zdi_maps}. \textit{Bottom panel:} Fit of the light curve. In the first plot, we show the 741 relative photometry values from TESS data in black and the ZDI fit in red. The second plot displays the residuals between the data and the fit, exhibiting a RMS dispersion of 2.4~mmag. }

    \label{fig:ZDI_reconstruction_TESS}
\end{figure}

In a second step, we reconstructed the brightness distribution and large-scale magnetic topology from the Stokes~$I$ and $V$ LSD profiles and the 741 additional photometric data points, proceeding as described in \cite{finociety21} to take into account (with the Planck function) that the spectropolarimetric and photometric data have different average wavelengths. Both spectroscopic and photometric data were fitted down to a unit $\chi^2_r$. We find that the spot coverage is increased up to 9\% and the reconstructed map features more low-latitude structures and enhanced contrasts between the reconstructed spots and the quiet photosphere (top panel of Fig.~\ref{fig:ZDI_reconstruction_TESS}). We however see no major differences in the location of the reconstructed brightness features. The residuals in the fitted light curve show low-amplitude structures that were not fitted (bottom panel of Fig.~\ref{fig:ZDI_reconstruction_TESS}), likely due to, rapidly evolving, surface features. Adding the TESS data in the fitting process does not change the magnetic maps as the photometry mainly informs on the contrast of the brightness features.

The magnetic topology is rather simple and similar to that obtained by \cite{donati14} from optical data.  
The average magnetic strength is equal to $\unsim1.9$~kG but can be locally more intense due to a strong radial field reaching up to $\unsim2.3$~kG. 
The poloidal component of the magnetic field, enclosing $\unsim65$\% of the overall reconstructed magnetic energy, mainly consists of a strong and axisymmetric dipole ($\unsim2.2$~kG) slightly tilted with respect to the rotation axis by $3^\circ$ towards phase 0.31, concentrating about 85\% of the poloidal energy.
The toroidal component is also mainly axisymmetric and characterized by a strong azimuthal field ring ($\unsim1.4$~kG) encircling the star close to the equator.
We also note that the regions associated with the strongest radial field coincide more or less with the largest cool spot.

In order to investigate whether the strength of the local magnetic field may also partly explain the variations of the EW of the Stokes~$I$ LSD profiles, we attempted to reconstruct the brightness and large-scale magnetic field of LkCa~4 using the filling factors $f_I$ and $f_V$ describing the fraction of the grid cells covered by small- and large-scale magnetic fields and respectively affecting Stokes~$I$ and Stokes~$V$ profiles, as defined in \cite{morin08}. More specifically, $B/f_V$ is the local strength of the magnetic field, $f_V$ the fraction of the cells contributing to the large-scale field, and $f_I$ the fraction of the cells contributing to the small-scale field. Whereas we set $f_V=f_I=1$ for the previous reconstructions\footnote{Line broadening being dominated by rotation, magnetic broadening has a minor impact on the Stokes I LSD profiles, hence our initial assumption on the $f_I$ and $f_V$ filling factors.}, we now assume $f_V=0.5$ and $f_I=1$, i.e., the local field is twice as strong as before, with only half of the cells contributing to the large-scale field and the whole cells to the small scale field. With these assumptions, we empirically find $\delta=0.9$ (compatible with the previous estimate), indicating that the observed variations of EW are indeed mostly due to the temperature of surface features and only slightly to the magnetic field. Besides, the reconstructed brightness and magnetic maps are globally consistent with the ones shown in Fig.~\ref{fig:zdi_maps}, with slightly less contrasted brightness features and a slightly weaker magnetic field. 

\begin{figure}
    \centering \hspace*{-0.2cm}
    \includegraphics[scale=0.27,trim={0cm .5cm 1cm 1cm},clip]{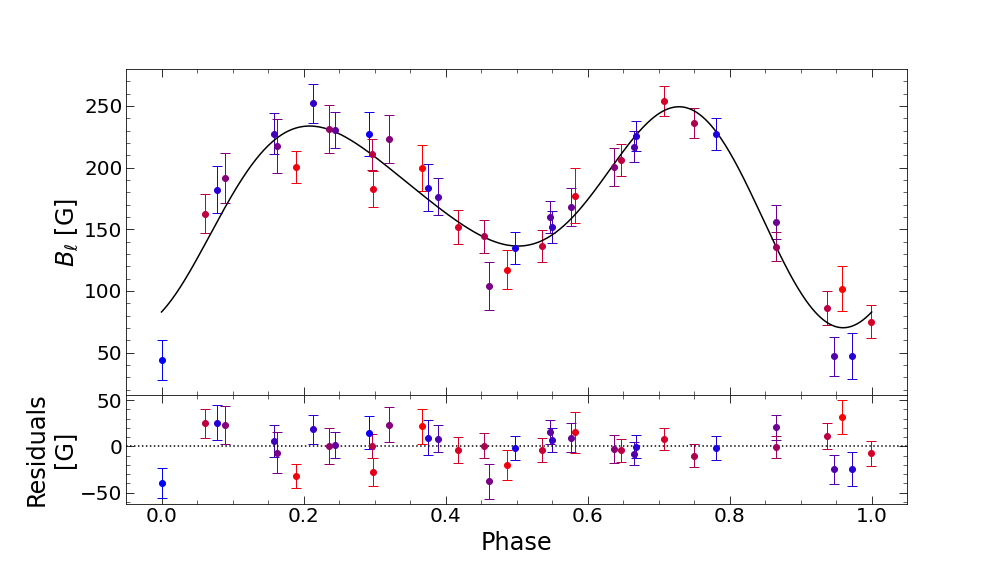}
    \caption{Phase-folded longitudinal field of LkCa~4. \textit{Top panel}: the measurements are displayed as coloured dots while the periodic fit is shown in solid black line. \textit{Bottom panel}: Residuals between the measurements and the model, exhibiting a RMS dispersion of 17~G corresponding to $\chi^2_r=1.14$. In both panels symbol colour depicts time, ranging from blue to red from the beginning to the end of our observations.}
    \label{fig:longitudinal_field}
\end{figure}

We computed the longitudinal field $B_\ell$ as the first moment of the Stokes~$V$ LSD profiles \citep{donati97}. We find that $B_\ell$ ranges from $\unsim50$ to $\unsim250$~G with uncertainties ranging from 12 to 23~G (median of 15~G). The $B_\ell$ measurements show a clear modulation and can be fitted down to the noise level with a periodic signal (sine + 2 harmonics) as no significant evolution is seen in these data (Fig.~\ref{fig:longitudinal_field}). This fit provides a stellar rotation period of $3.373\pm0.001$~d, consistent with TESS photometric variations ($3.372\pm0.002$~d) and with the value provided when using a quasi-periodic GP to model the longitudinal field ($3.373\pm0.002$~d). 

\begin{figure*}
    \centering
    
    \includegraphics[scale=0.16,trim={1cm 8cm 1cm 1cm},clip]{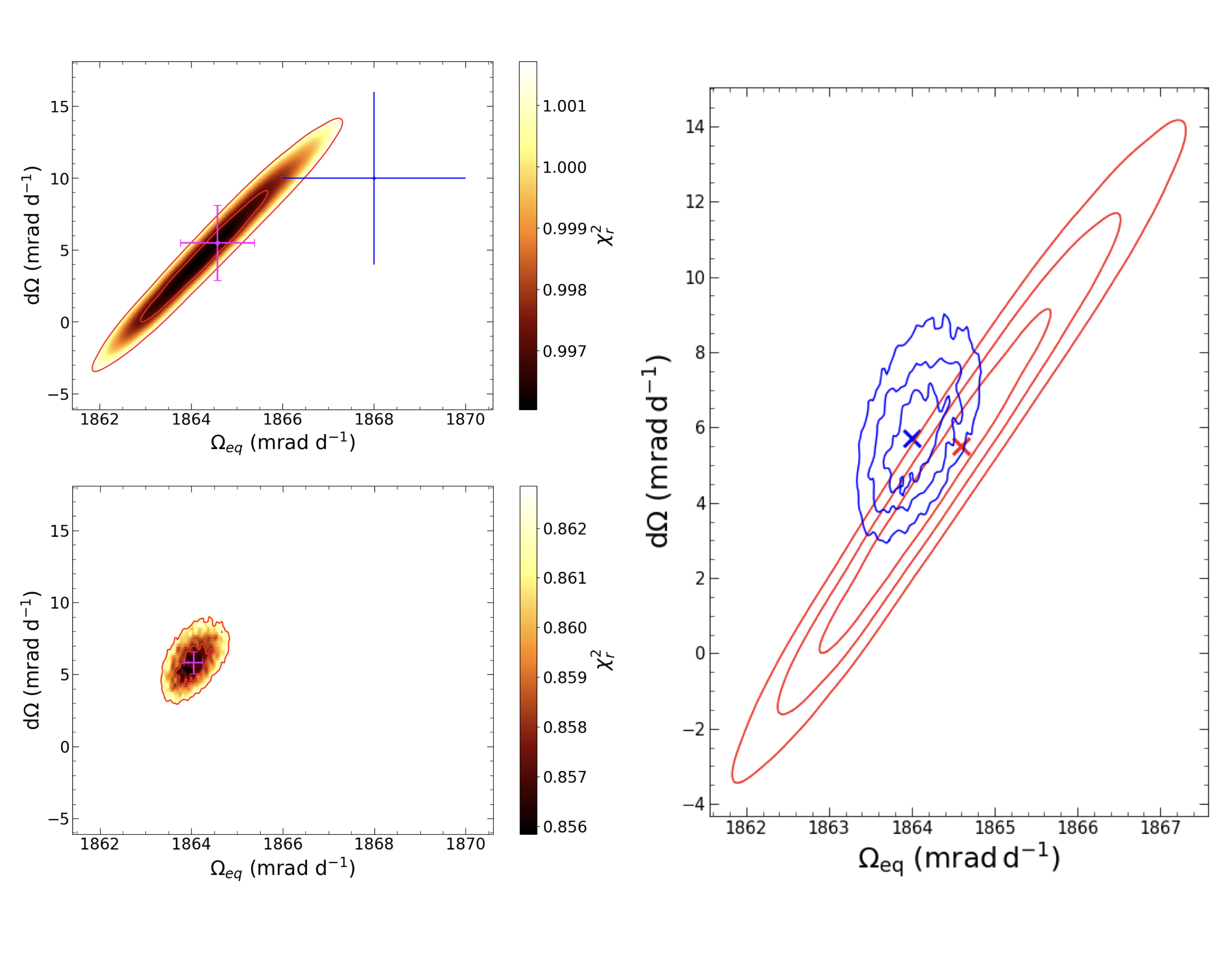}

    \caption{Differential rotation at the surface of LkCa~4 as measured from SPIRou data. \textit{Left panels:} $\chi^2_r$ maps in the \{$\Omega_{\rm eq}$, d$\Omega$\} space obtained from Stokes~$I$ (top) and $V$ (bottom) LSD profiles, with $\Omega_{\rm eq}$ and d$\Omega$ being the rotation rate at the equator and the difference of rotation rate between the pole and the equator, respectively. Red ellipses show the contours of 68\% (1$\sigma$) and 99.7\% (3$\sigma$) confidence levels for both parameters as a pair. The magenta cross depicts our optimal value with its associated error bars derived from a paraboloid fit to the $\chi^2_r$ map, while the blue one in the top plot corresponds to the estimate reported by \citet{donati14} from optical data. \textit{Right panel:} Comparison of the contours associated with the 1$\sigma$, 2$\sigma$ and 3$\sigma$ confidence levels obtained from Stokes~$I$ (red) and $V$ (blue) LSD profiles. The coloured crosses depicts the optimal values derived with SPIRou data. }
    \label{fig:differential_rotation}
\end{figure*}

\subsection{Surface differential rotation}
\label{sec:differential_rotation}
Our data are spread over 32 rotation cycles, during which the brightness map and the large-scale magnetic topology have potentially evolved under the effect of differential rotation, although the expected level of differential rotation is weak \citep{donati14}. ZDI allows one to take into account this variability assuming that the rotation rate at the surface of the star follows a solar-like law given by :

\begin{equation}
    \Omega(\theta) = \Omega_{\rm eq} - \left(\cos{\theta}\right)^2\,\mathrm{d}\Omega
\end{equation}

where $\theta$ is the colatitude, $\Omega_{\rm eq}$ and d$\Omega$ the parameters of the differential rotation law, characterizing the rotation rate at the equator and the pole-to-equator rotation rate difference, respectively.

In practice, we set the amount of information that ZDI is allowed to reconstruct and we look for the DR parameters that minimize $\chi^2_r$. Fitting Stokes~$I$ and $V$ profiles independently yields $\chi^2_r$ maps from which we derive optimal parameters and their error bars (\citealt{donati03}; see Fig.~\ref{fig:differential_rotation}). From the brightness map we derive $\Omega_{\rm eq}=1864.6\pm0.8$~$\mathrm{mrad\,d^{-1}}$ and d$\Omega=5.5\pm2.6$~$\mathrm{mrad\,d^{-1}}$ (implying a period of $3.3697\pm0.0014$ and $3.3797\pm0.0049$~d at the equator and at the pole, respectively). 
The magnetic maps yield fully compatible but more accurate values, with error bars more than $3\times$ smaller. We find $\Omega_{\rm eq}=1864.0\pm0.2$~$\mathrm{mrad\,d^{-1}}$ and d$\Omega=5.8\pm0.8$~$\mathrm{mrad\,d^{-1}}$ (corresponding to a period of $3.3708\pm0.0004$ and $3.3813\pm0.0015$~d at the equator and at the pole, respectively).
The slope of the major axis of the confidence ellipse provides an estimate of the colatitude associated with the barycentre of the brightness or magnetic distribution, found to be equal to $\unsim55^\circ$ and $\unsim60^\circ$. These results indicate that both brightness and magnetic regions are, in average, located at similar latitudes, explaining why the longitudinal field, the photometry and the RVs (Section~\ref{sec:radial_velocities}) yield similar values for the stellar rotation period.

The estimates of differential rotation inferred from the Stokes~$I$ LSD profiles are sensitive to the value of $\delta$. Varying $\delta$ within the error bars (i.e. between 0.9 and 1.1) nevertheless yields estimates of d$\Omega$ that are still compatible with the value reported above within 1$\sigma$.

We chose a unique set of parameters to described the DR at the surface of LkCa~4, taken as the weighted means of both estimates: $\Omega_{\rm eq}=1864.0\pm0.2$~$\mathrm{mrad\,d^{-1}}$ and d$\Omega=5.8\pm0.8$~$\mathrm{mrad\,d^{-1}}$. These weighted means actually yield the same values as those derived from our Stokes~$V$ LSD profiles since they are more accurate than those provided by our Stokes~$I$ profiles. This implies that the rotation period ranges from $3.3708 \pm 0.0004$~d at the equator to  $3.3813 \pm 0.0015$~d at the pole. The surface DR of LkCa~4 is thus 10$\times$ weaker than that of the Sun and significantly different from 0 at a 7$\sigma$ level.

\section{Filtering the activity jitter}
\label{sec:radial_velocities}

To investigate the impact of activity on the RV curve (activity jitter), we computed the RV of the star corresponding to each spectropolarimetric observation as the first moment of the Stokes~$I$ LSD profiles (e.g. \citealt{donati17,yu19,finociety21}). We estimated the associated photon-noise uncertainties from the dispersion of the RV measurements on simulated noisy profiles (from the synthetic set of profiles provided by ZDI) featuring the same SNR as the observed ones for several realisations of noise. The computed error bars are typically equal to 0.15~\kms, thus $2.5\times$ larger than those obtained from ESPaDOnS data \citep{donati14}. Such a difference is most likely related to the depth of the spectral lines, about $3.5 \times$ shallower in the infrared than in the optical. The measured RVs exhibit a RMS dispersion of 1.45~\kms\ with a $\chi^2_r=84.89$, with respect to a model with constant RV.

\begin{figure*}
    \centering \hspace*{-0.5cm}
    \includegraphics[scale=0.4,trim={1cm 1cm 1cm 1cm},clip]{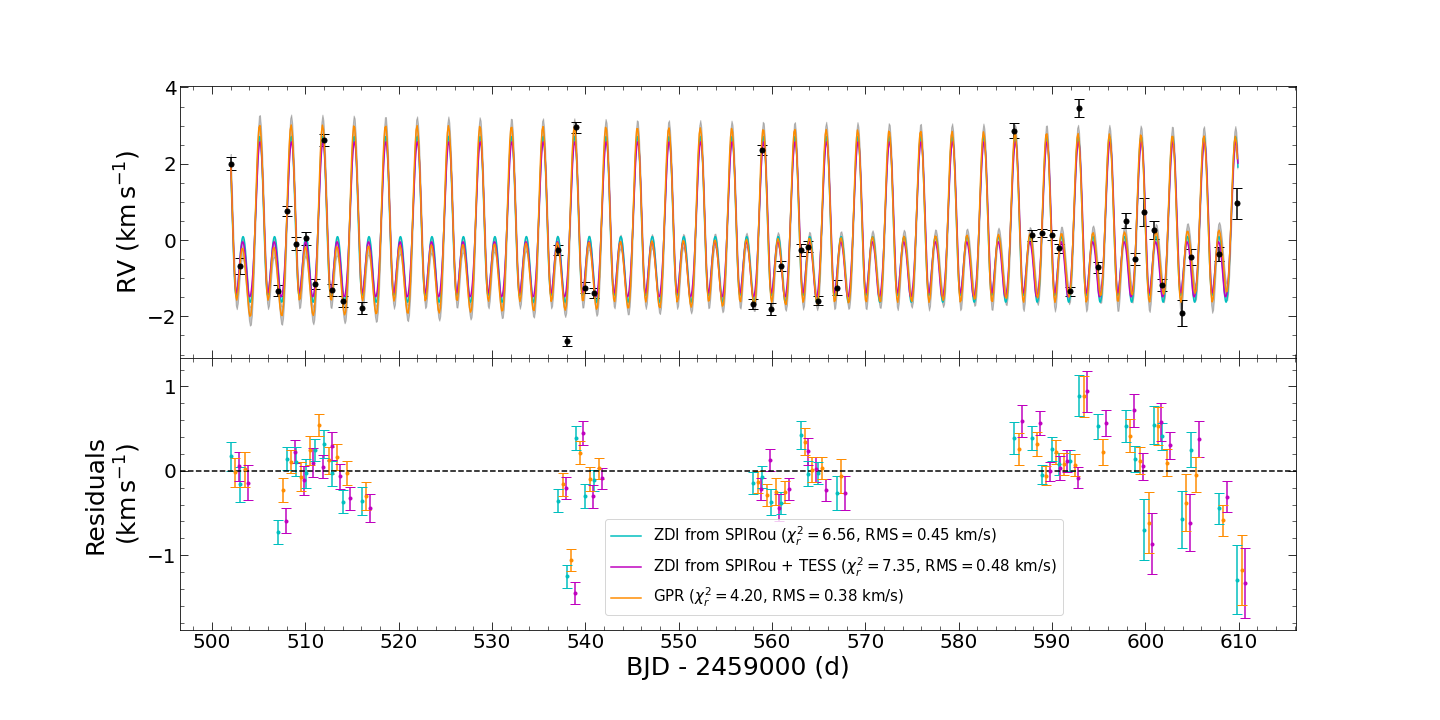}
    \caption{RVs of LkCa~4 measured from SPIRou data. \textit{Top panel}: The raw observed RVs are shown as black dots with their associated error bars. They exhibit a full amplitude of 6.10~\kms\ and a RMS dispersion of 1.46~\kms. The cyan and magenta curves correspond to models derived from the ZDI reconstruction taking into account SPIRou data alone or SPIRou and TESS data simultaneously, respectively. The orange curve represents the GPR with the 1$\sigma$ confidence area in light grey. \textit{Bottom panel}: Filtered RVs associated to each of the 3 models, with the same colour code as in the top panel. We obtained RMS dispersions of the filtered RVs of 0.45, 0.48 and 0.38~\kms, for cyan, magenta and orange models, respectively. We slightly shift the filtered RVs along the horizontal axis for each model for clarity purposes.}
    \label{fig:radial_velocities}
\end{figure*}

We computed the RVs associated with the synthetic Stokes~$I$ profiles provided by the reconstructed brightness maps obtained with ZDI. Comparing these values with the raw measurements, we see that the RVs are fitted down to $\chi^2_r=6.56$ when applying ZDI to SPIRou data alone and down to $\chi^2_r=7.35$ when ZDI is applied to SPIRou and TESS data simultaneously. The corresponding RMS dispersion of the filtered RVs (i.e. the difference between the observed and modeled RVs) is equal to 0.45~\kms\ and 0.48~\kms, showing that adding photometry in our process only slightly degrades our jitter filtering efficiency (see Fig.~\ref{fig:radial_velocities}).

We also used a GP with the quasi-periodic kernel introduced in Sec~\ref{sec:tess}. The log likelihood function we maximize (Eq.~\eqref{eq:log_likelihood}) takes into account an additional term corresponding to an excess of uncorrelated noise $s$ as in \cite{finociety21}.

\begin{equation}
    \log \mathcal{L} = -\frac{1}{2}\left(N_0 \log{2\pi} + \log{|\boldsymbol{\mathsf{K}}+\boldsymbol{{\Sigma}}+\boldsymbol{\mathsf{S}}| + \boldsymbol{y}^T(\boldsymbol{\mathsf{K}}+\boldsymbol{{\Sigma}}+\boldsymbol{\mathsf{S}})^{-1}\boldsymbol{y}}  \right) 
    \label{eq:log_likelihood}
\end{equation}

where $N_0$ refers to the number of measured RVs, $\boldsymbol{\mathsf{K}}$ is the covariance matrix associated to the quasi-periodic kernel, $\boldsymbol{{\Sigma}}$ is the diagonal matrix containing the squared error bars on the measurements, $\boldsymbol{\mathsf{S}}=s^2\,\boldsymbol{\mathsf{I}}$ with $\boldsymbol{\mathsf{I}}$ being the identity matrix and $\boldsymbol{y}$ is the vector gathering the measured RVs.

We sampled the posterior distributions of all the parameters (i.e. the amplitude of the GP $\theta_1$, the exponential decay timescale $\theta_2$, the rotation period $\theta_3$, the smoothing parameter $\theta_4$ and the excess of uncorrelated noise $s$) with a Monte Carlo Markov Chain (MCMC) approach using the \textsc{emcee python} module \citep{emcee}. More specifically, we ran our MCMC on 5000 iterations of 100 walkers and then removed a burn-in period of 500 iterations before estimating the best value of each parameter, chosen to be the median of the posterior distributions. 

\begin{table}
	
	\caption{Best value for the parameters involved in the GPR derived from the MCMC approach. We list the hyperparameters in the first column, the priors used in the MCMC in the second column (with the lower and upper limits of the interval) and the best estimate (taken as the median of the posterior distribution) in the last column.}
	\label{tab:parameters_GPR}
	\centering

	\begin{tabular}{lcc}
	\hline
	
	Hyperparameter & Prior & Estimate \\ \hline
	
	$\theta_1$ [$\mathrm{km\,s^{-1}}$] & Uniform (0, 10) & $1.87\pm0.86$ \\
	$\theta_2$ [$\rm d$] & Uniform (0, 400)  & $262\pm85$ \\
	$\theta_3$ [d] & Uniform (3,4) & $3.372\pm0.002$ \\
	$\theta_4$  & Uniform (0,1) & $0.50\pm0.14$\\
	$s$ [$\mathrm{km\,s^{-1}}$] & Uniform (0,1) &$0.37\pm0.06$  \\ \hline
	\end{tabular}
	
\end{table}

The optimal GP parameters are listed in Table~\ref{tab:parameters_GPR} whereas the best fit is shown in Fig~\ref{fig:radial_velocities}. GPR yields a 15\% lower dispersion of the filtered RVs with respect to ZDI of 0.38~\kms\, associated with a $\chi^2_r=4.20$. In any case, the dispersion of the filtered RVs is 2.5--3 times larger than the typical measured error bars (0.15~\kms).

\begin{figure}
	\includegraphics[scale=0.275]{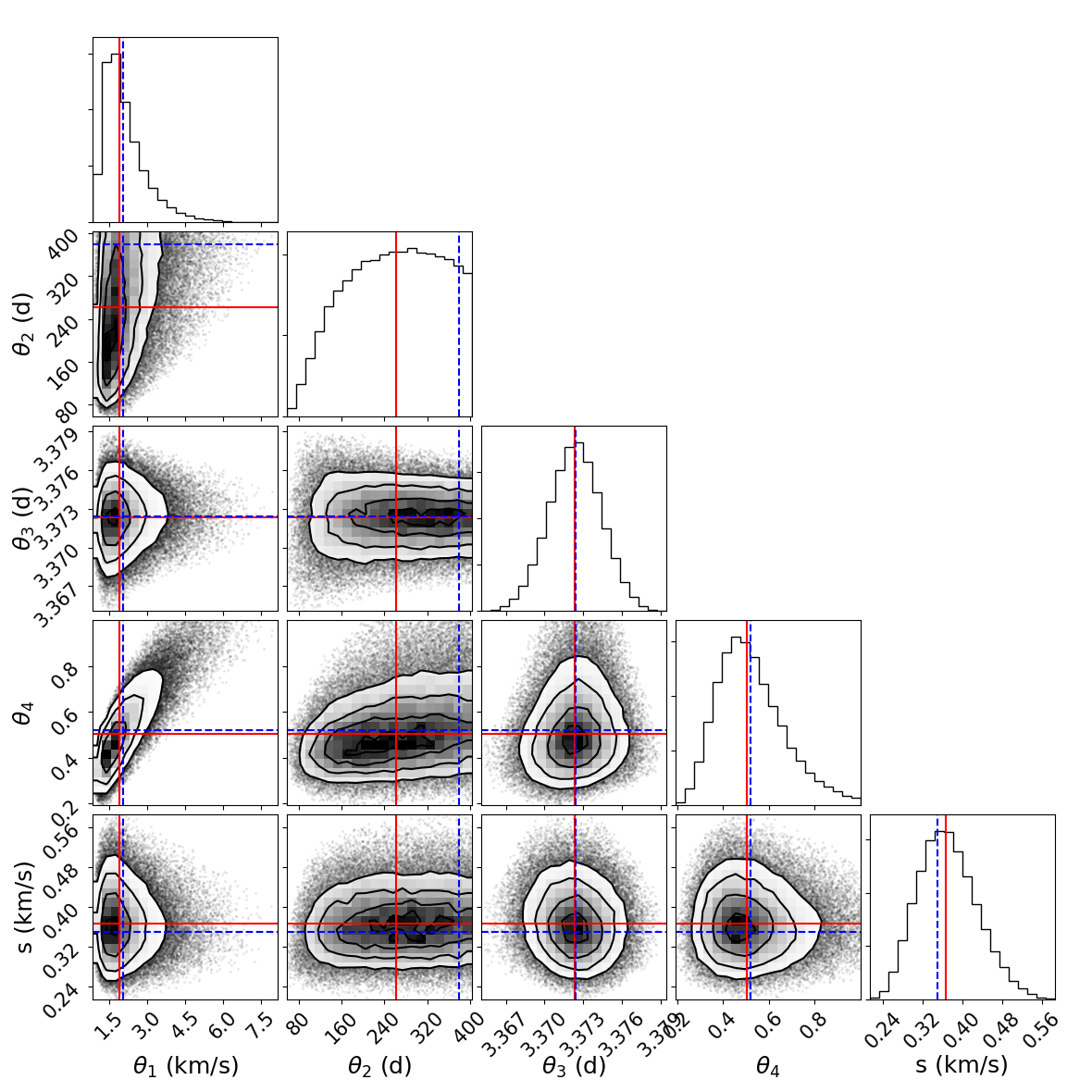}
	\caption{Corner plot of the posterior distribution of the parameters of the GPR given by the MCMC sampling. The median values of the posterior distributions are shown as red solid lines and correspond to the best estimates listed in Table~\ref{tab:parameters_GPR}. The blue dashed lines depict the values that maximize the log likelihood. This plot was generated with the \textsc{corner python} module \citep{corner}.}
	\label{fig:corner_plot_rv}
\end{figure}

From our MCMC, we note that $s=0.37\pm0.06$~\kms\ is not compatible with 0, reflecting that our error bars are apparently underestimated, probably as a result of intrinsic variability induced by activity. In addition, the exponential decay-time scale is not well constrained (see corner plot in Fig.~\ref{fig:corner_plot_rv}) indicating that the rotationally modulated component of the activity signal remains stable over the time span of our observations, despite some dispersion observed at the end of the SPIRou campaign (most likely due to bad weather conditions in 2022 January). The stellar rotation period derived from these RVs is also equal to $3.372\pm0.002$, again consistent with those inferred from TESS data and $B_\ell$ measurements. 

\begin{figure}
    \centering
    \hspace*{-.65cm}
    \includegraphics[scale=0.245, trim= 0cm 3cm 3cm 3 cm]{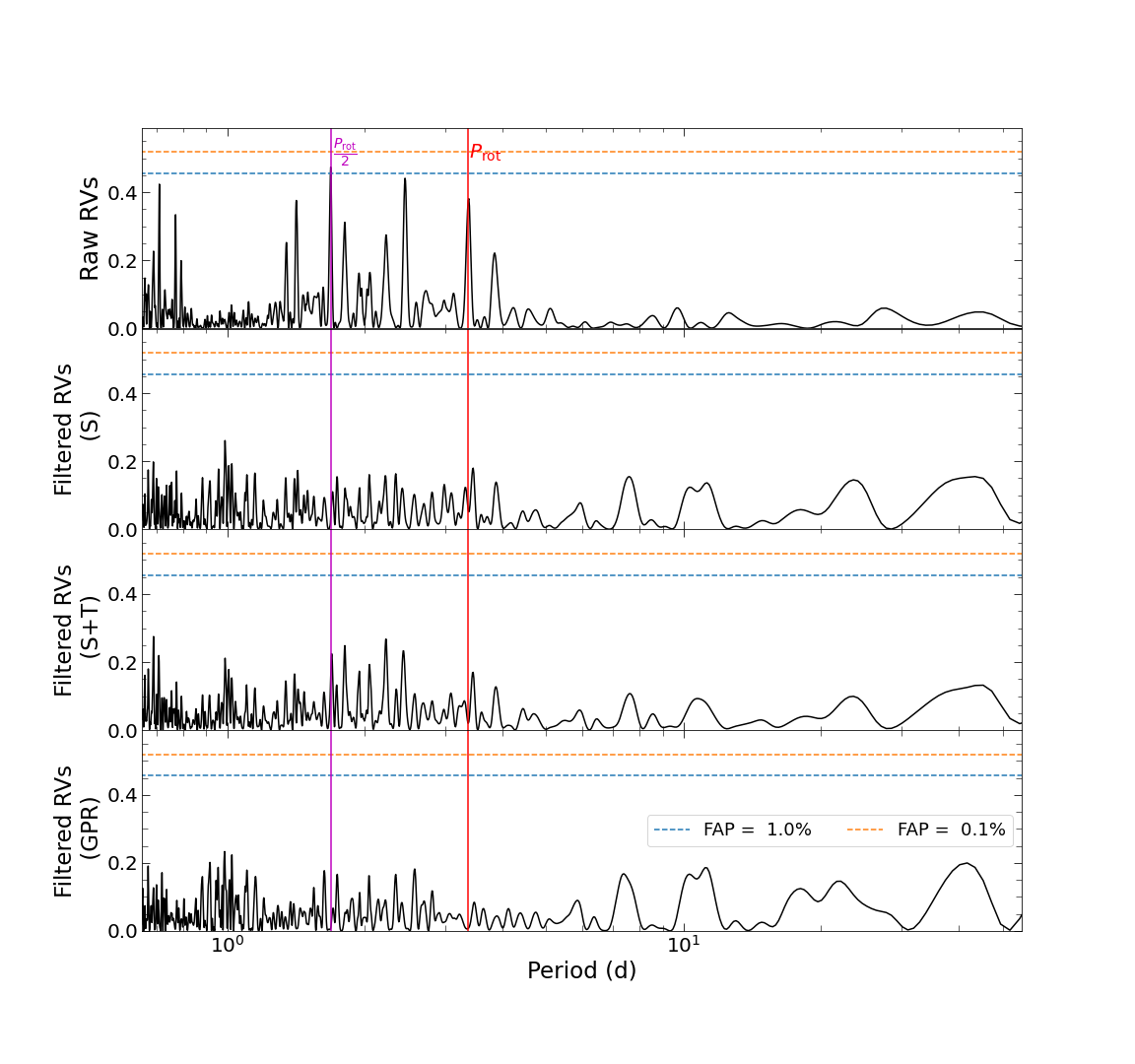}
    \caption{Periodograms of raw radial velocities ($1^{\rm st}$ panel) and filtered radial velocities computed from ZDI brightness reconstructions when using SPIRou (S) data only ($2^{\rm nd}$ panel) or SPIRou and TESS (S+T) data simultaneously ($3^{\rm rd}$ panel) or GPR ($4^{\th}$ panel). The red and magenta vertical lines depict $P_{\rm rot}$ and $P_{\rm rot}/2$. The horizontal dashed lines indicate the FAP levels at 1\% and 0.1\%. These periodograms were generated using the \textsc{pyastronomy python} module \citep{pyastronomy}.}
    \label{fig:periodograms_RV}
\end{figure}

We computed the Generalised Lomb-Scargle Periodogram \citep{zechmeister09} using the \textsc{pyastronomy python} module \citep{pyastronomy} of the raw and filtered RVs obtained with the three models (see Fig.~\ref{fig:periodograms_RV}). We do not see any significant peak in these periodograms, indicating that our data do not provide any evidence for a hJ orbiting LkCa~4.
In order to investigate the upper mass limit of a potential planet that can be detected from our data, we proceeded as in \cite{yu19} and \cite{finociety21}. We first simulated RV curves including the RV activity jitter (using the derived GPR) and a planetary signature (assuming a circular orbit for various masses and distances from the host star). For each curve, we collected 41 measurements following the same temporal sampling as our actual SPIRou data, for which we added a white noise of 0.400~\kms\ (i.e. taking into account the photon-noise uncertainty of our observations and the uncorrelated noise derived from our GPR). We then fitted the simulated datasets using (i) a model including the activity jitter only and (ii) a model including both contributions of the activity jitter and the planet. We assume that a planet is reliably detected (i.e. at a $>3\sigma$ level) when the difference in logarithmic marginal likelihood is larger than 10 (e.g. \citealt{yu19,finociety21}). From these simulations, we find that we can safely claim that a planet is detected only if the semi-amplitude of the induced RV signal is larger than 0.44~\kms, which corresponds to a 4.3~M$_{\rm jup}$ planet at 0.1~au (i.e. $P=13$~d). 

\section{Activity indicators}
\label{sec:activity}
\subsection{Equivalent width variations}

We focussed on the \ion{He}{i} triplet (1083~nm; Fig.~\ref{fig:spectral_line_he}), the Paschen beta (Pa$\beta$; 1282~nm; Fig.~\ref{fig:spectral_line_pa}) and Brackett gamma (Br$\gamma$; 2165~nm; Fig.~\ref{fig:spectral_line_br}) lines, known to be proxies of activity in the NIR (e.g. \citealt{zirin82,short98}). We see that the Pa$\beta$ line is blended with a Ca line, which creates a bump in the blue wing of Pa$\beta$. However, this feature apparently does not vary more than the continuum and would therefore not impact our analyses.

We computed the equivalent width variations (EWVs) of each line, quantifying the changes in the EW of these lines due to stellar activity, in a way similar to that of \cite{finociety21}. With this definition, a negative EWV indicates enhanced absorption with respect to the median spectrum. In practice, we first divided each telluric-corrected Stokes~$I$ spectrum (top panel of Figs.~\ref{fig:spectral_line_he}, \ref{fig:spectral_line_pa} and \ref{fig:spectral_line_br}) by the median one (bottom panel of Figs.~\ref{fig:spectral_line_he}, \ref{fig:spectral_line_pa} and \ref{fig:spectral_line_br}) in the stellar rest frame, then computed the EW of the median-divided spectra (middle panel of Figs.~\ref{fig:spectral_line_he}, \ref{fig:spectral_line_pa} and \ref{fig:spectral_line_br}), by fitting a Gaussian function of full-width-at-half-maximum set to 55~\kms (consistent with the median profile of the lines) centred on the stellar rest frame. 

Assuming equal error bars for all spectral points from the dispersion between spectra in the continuum, we found that the average photon-noise uncertainties on the EWVs are equal to 0.7, 0.8 and 2.6~pm for the \ion{He}{I}, Pa$\beta$ and Br$\gamma$ lines, respectively. As in the case of V410~Tau \citep{finociety21}, these error bars are actually underestimated as they do not account for intrinsic variability, e.g., like that induced by activity. We therefore scaled-up all these error bars to ensure a unit $\chi^2_r$ between a periodic fit and the measurements, yielding empirical (and likely pessimist) error bars taking into account photon-noise and intrinsic variability. In practice, the \ion{He}{I} EWVs were fitted with a pure sine wave, while the Pa$\beta$ EWVs were better modeled with a sine wave including the first harmonic. The Br$\gamma$ EWVs show no significant variation and are compatible with 0, meaning that no rotationally modulated activity signal is detected in this line. The enhanced uncertainties we obtain are equal to 5.4, 1.9 and 6.6~pm for the \ion{He}{I}, Pa$\beta$ and Br$\gamma$ EWVs, respectively, i.e., 1.7, 2.4 and 2.5 times larger than the average photon noise ones quoted above.  

\begin{figure}
    \centering
    \hspace*{-0.4cm}
    \begin{subfigure}{0.49\textwidth}
         \centering
         \includegraphics[scale=0.25,trim={0cm 0cm 3.5cm 2cm},clip]{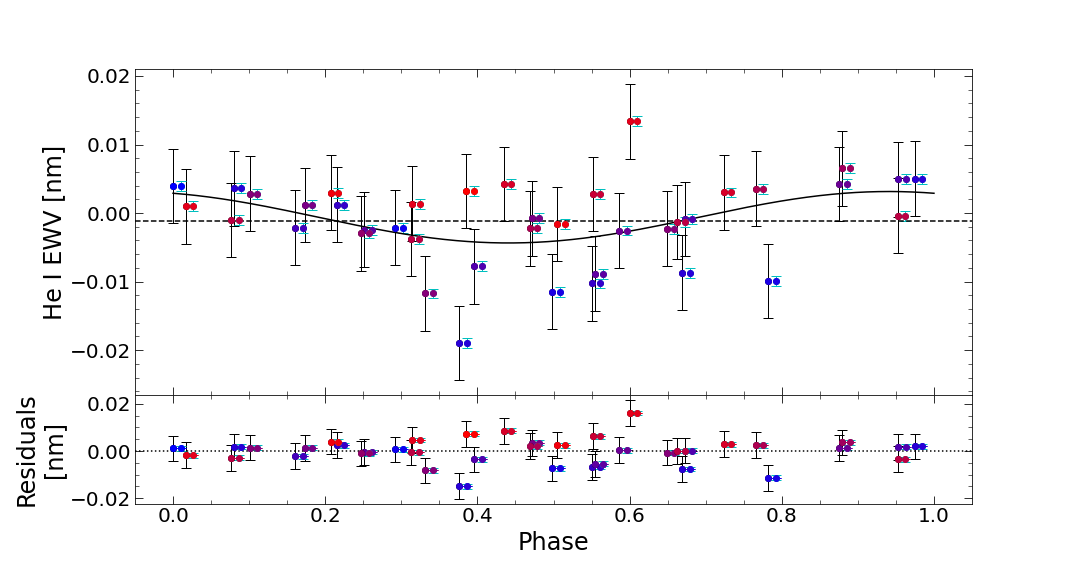}
         
    \end{subfigure}
    \hfill
    \hspace*{-0.4cm}
    \begin{subfigure}{0.49\textwidth}
         \centering
         \includegraphics[scale=0.25,trim={0cm 0cm 3.5cm 2cm},clip]{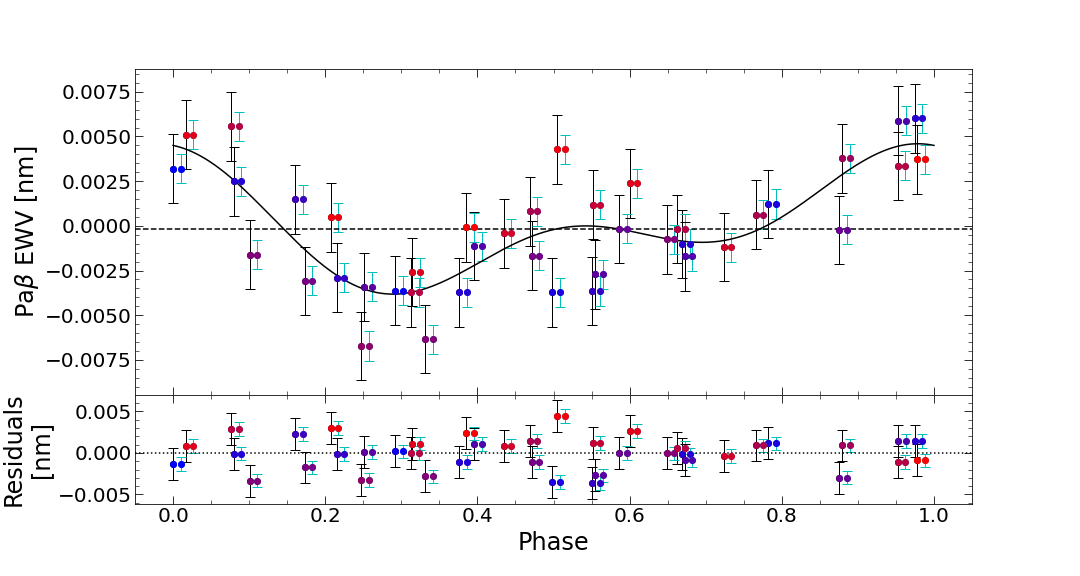}
         
    \end{subfigure}
    \hfill
    \hspace*{-0.4cm}
    \begin{subfigure}{0.49\textwidth}
         \centering
         \includegraphics[scale=0.25,trim={0cm 0cm 3.5cm 1cm},clip]{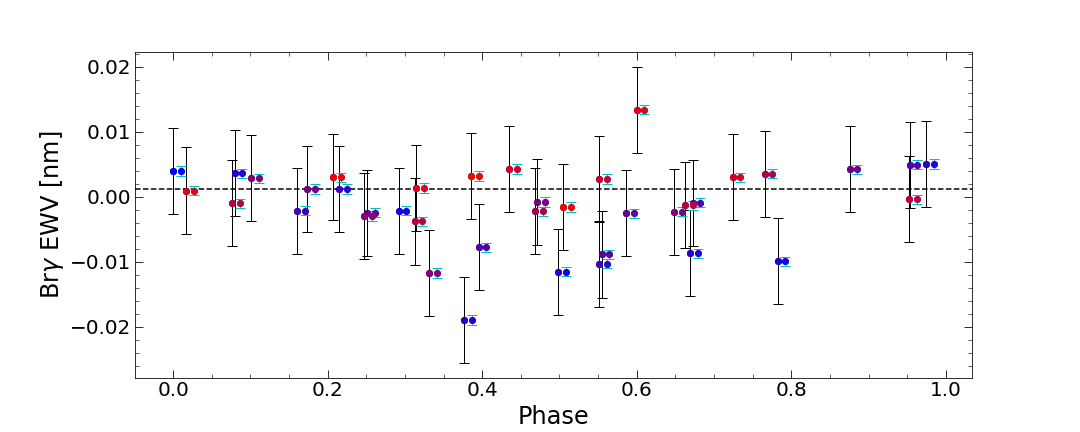}
         
    \end{subfigure}
    
    \caption{Phase folded activity EWVs derived from the \ion{He}{i} triplet at 1083.3~nm (first panel), Pa$\beta$ (second panel) and Br$\gamma$ lines (third panel). For the \ion{He}{i} and Pa$\beta$ lines, we show the fit to the data with a periodic function as a solid black line and with a constant as a dotted line in the top plot while the residuals are shown in the bottom plot. For Br$\gamma$ we only show the fitted constant compatible with 0. In all panels, the cyan error bars are those expected from photon noise (0.7, 0.8 and 2.6~pm) while the black ones are those ensuring a unit $\chi^2_r$ fit to the data (5.4, 1.9 and 6.6~pm). The cyan error bars are slightly shifted along the horizontal axis for display purposes. The colour of the dots traces the rotation cycle going from blue (first observation) to red (last observation).}

    \label{fig:activity_proxies}
\end{figure}

From the empirical error bars, we can estimate a false alarm probability for the detected modulation. In practice, we fitted a constant instead of a periodic signal to the \ion{He}{I} and Pa$\beta$ EWVs, and we used the empirical error bars to compute the associated $\chi^2_r$. We find $\chi^2_r=1.23$ and 2.80, for the \ion{He}{i} and Pa$\beta$ EWVs, respectively, corresponding to a probability for the detected modulation to be spurious by chance of 0.15 and $10^{-8}$.
This indicates that we detected a significant modulation in the Pa$\beta$ line that shows enhanced absorption around phases 0.3 and 0.7. The \ion{He}{I} EWVs show a much less significant modulation, with enhanced absorption around phase 0.4 (see Fig.~\ref{fig:activity_proxies}). 

\subsection{2D Periodograms}

We computed 2D periodograms for the \ion{He}{i}, Pa$\beta$ and Br$\gamma$ lines. We proceeded as in \cite{finociety21}, i.e. we computed a Generalised Lomb-Scargle Periodogram normalized to 1 following \cite{zechmeister09}, for each velocity bin of the median-divided spectra between $-100$ and $+100$~\kms, thanks to the \textsc{pyastronomy python} module \citep{pyastronomy}. With this normalization, a value of 1 indicates a perfect sinusoidal fit to the data.

\begin{figure}
    \centering
    \hspace*{-0.4cm}
    \begin{subfigure}{0.49\textwidth}
         \centering
         \includegraphics[scale=0.25,trim={0cm 3cm 3cm 5cm},clip]{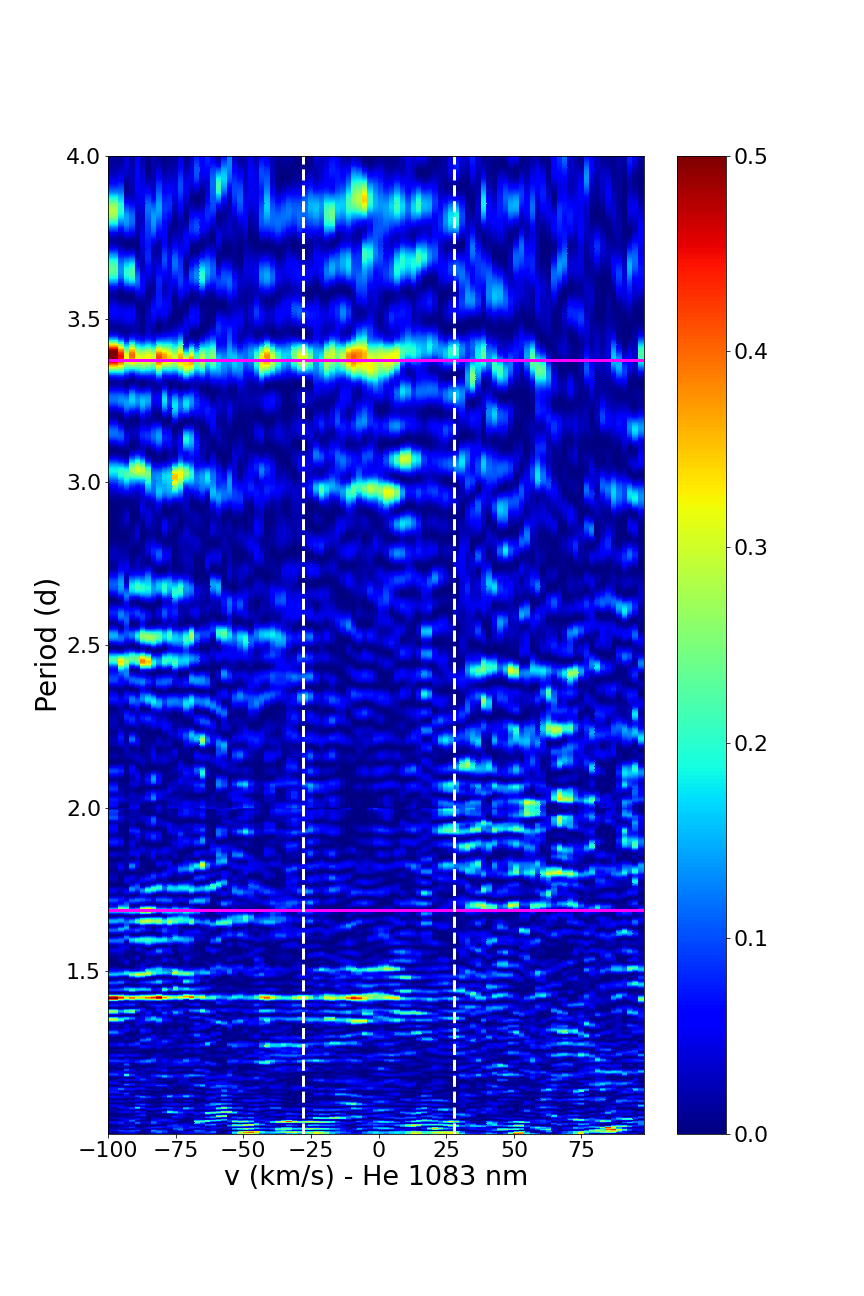}
    \end{subfigure}
    \hfill
    \hspace*{-0.4cm}
    \begin{subfigure}{0.49\textwidth}
         \centering
         \includegraphics[scale=0.25,trim={0cm 3cm 3cm 5cm},clip]{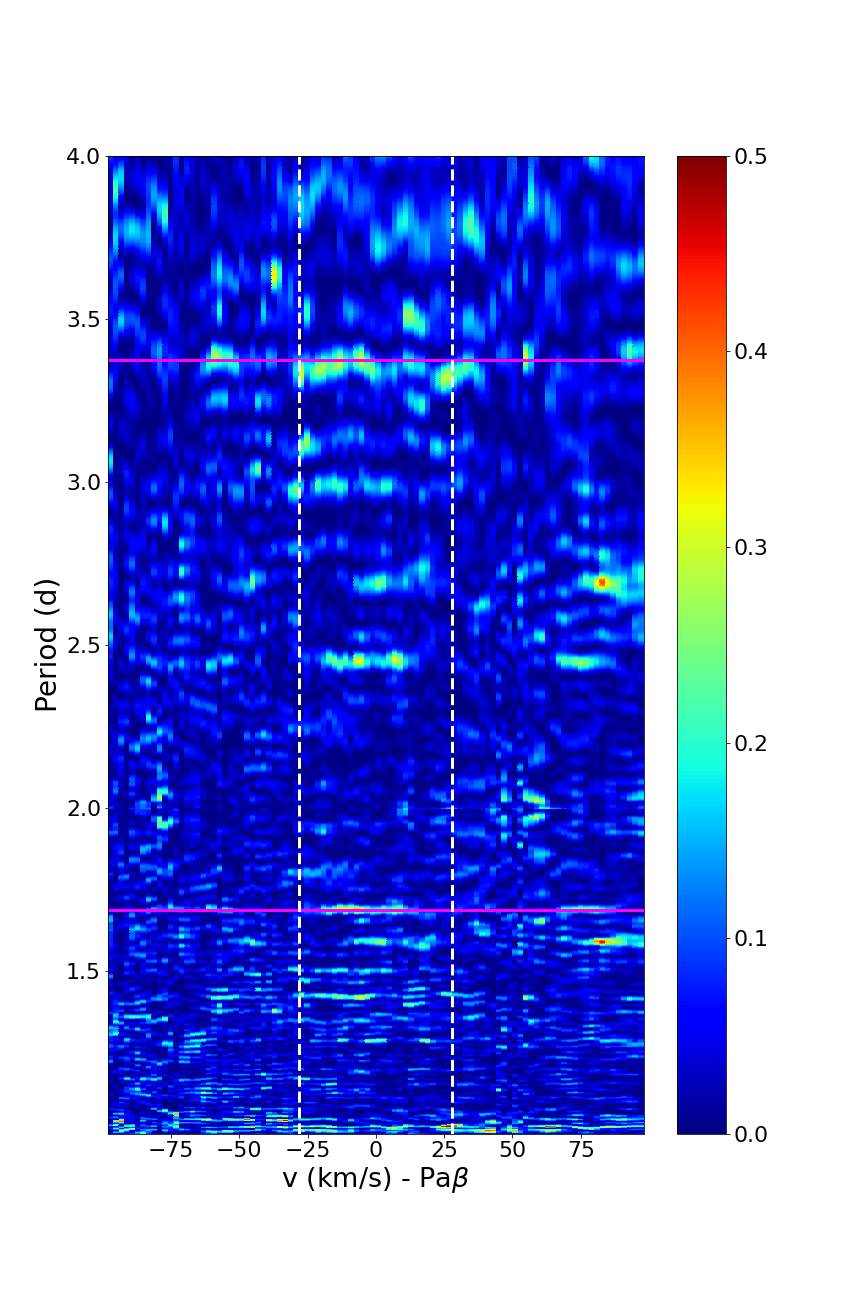}
    \end{subfigure}
    %\hfill
    %\hspace*{-0.4cm}
    %\begin{subfigure}{0.49\textwidth}
    %     \centering
    %     \includegraphics[scale=0.25,trim={0cm 0cm 3.5cm 2cm},clip]{Fig_Sec5/Ga_periodogram.png}
         
    %\end{subfigure}

    \caption{2D Periodograms of the \ion{He}{i} (top) and Pa$\beta$ (bottom) lines computed with the \textsc{pyastronomy python} module \citep{pyastronomy}. The magenta line depicts the stellar rotation period and half the rotation period. In both panels, the dashed vertical white lines represents $\pm v\sin{i}$. The colour bar reflects the power of the Generalized Lomb-Scargle periodogram associated with the period in each velocity bin, being normalized following \citet{zechmeister09}, i.e. with a value of 1 indicating a perfect fit to the data for the corresponding period. The colour bar is limited up to 0.5 for display purposes.}

    \label{fig:2D_periodograms}
\end{figure}

We show the results in Fig.~\ref{fig:2D_periodograms}, where we see a modulation with a period close to the stellar rotation period for the \ion{He}{i} triplet. For the Pa$\beta$ line, we see a peak in the periodogram at the stellar rotation period but also at half the rotation period (see Fig.~\ref{fig:2D_periodograms}). 
The Br$\gamma$ 2D periodogram does not show any modulation even at half the rotation period. We also see a signal at the stellar rotation period around $-85$~\kms\ in the \ion{He}{i} periodogram, which is most likely related to the modulation of a nearby photospheric Ca line.

These results are consistent with the detected modulation in the EWVs for the \ion{He}{i} and Pa$\beta$ lines. In particular, we see that only one period shows up in the \ion{He}{i} periodogram (the associated EWVs being fitted with a pure sine curve) while the Pa$\beta$ EWVs are fitted with a slightly more complex curve including the fundamental and the first harmonic, with both showing up in the 2D periodogram.

\subsection{Autocorrelation matrices}

\begin{figure}
    \centering
    \hspace*{-0.4cm}
    \begin{subfigure}{0.49\textwidth}
         \centering
         \includegraphics[scale=0.25,trim={0cm 0cm 3.cm 2cm},clip]{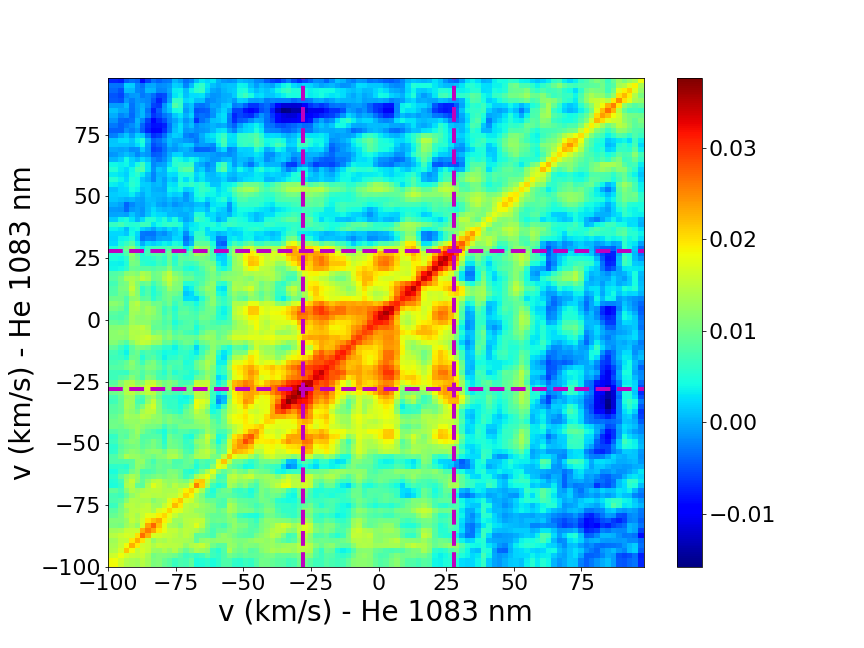}
         
    \end{subfigure}
    \hfill
    \hspace*{-0.4cm}
    \begin{subfigure}{0.49\textwidth}
         \centering
         \includegraphics[scale=0.25,trim={0cm 0cm 3.cm 2cm},clip]{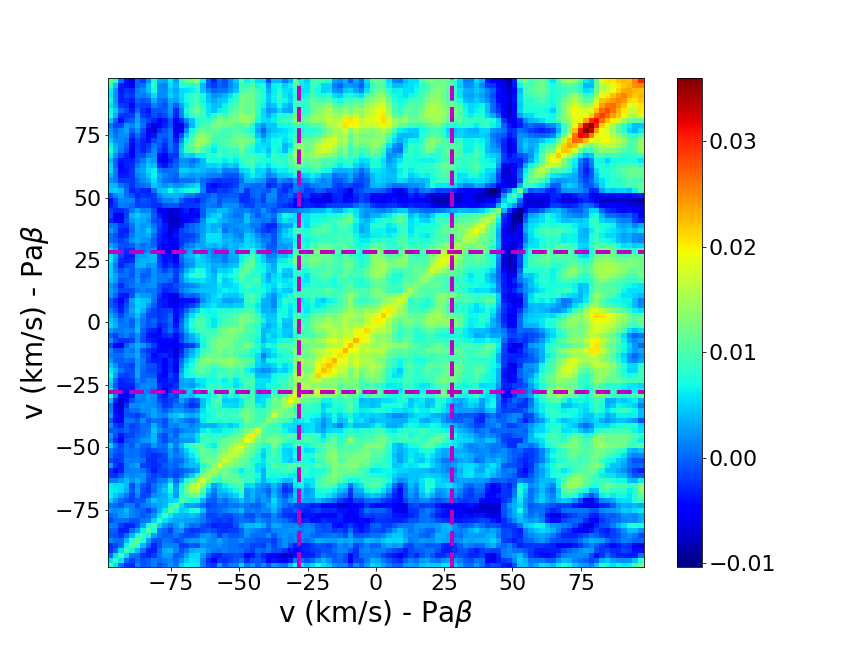}
         
    \end{subfigure}
    %\hfill
    %\hspace*{-0.4cm}
    %\begin{subfigure}{0.49\textwidth}
    %     \centering
    %     \includegraphics[scale=0.25,trim={0cm 0cm 3.cm 2cm},clip]{Fig_Sec5/Ga_Pearson_unnormalised2.png}
         
    %\end{subfigure}

    \caption{Autocorrelation matrices for \ion{He}{i} (top panel) and Pa$\beta$ (bottom panel). The colorbar refers to the value of the unnormalized coefficients as defined in \citet{finociety21} with the scale depending on the level of variability for each of the considered line. Correlations show up in reddish colours while anticorrelation are represented by blueish colours. The dashed magenta lines depicts $\pm v\sin{i}$.}

    \label{fig:autocorrelation_matrices}
\end{figure}

We proceeded as in \cite{finociety21} to compute the autocorrelation matrices of the three lines, within an interval of $\pm100$~\kms. In particular, using the definition of the unnormalized correlation coefficient given in \cite{finociety21} allows one to highlight the relative importance of the correlations. The \ion{He}{i} and Pa$\beta$ matrices are shown in Fig.~\ref{fig:autocorrelation_matrices}.

We see a clear autocorrelation of the \ion{He}{i} triplet and a less obvious one for the Pa$\beta$ line. The autocorrelation matrix of these two lines shows a slight asymmetry (as the 2D periodograms), the correlation being a bit more pronounced in the blue wing. 
As for V410~Tau, the Br$\gamma$ autocorrelation matrices reveals no specific pattern and reflects mainly noise \citep{finociety21}.

\section{Summary and conclusions}
\label{sec:discussion}

Our paper describes results derived from NIR spectropolarimetric data, collected with SPIRou from 2021 Oct 14 to 2022 Jan 30, and photometric observations obtained with TESS between 2021 Sep 16 and Nov 06 and with the ground-based AZT-11 telescope at CrAO between 2021 Oct 10 and 2022 Jan 27, for the wTTS LkCa~4.

\subsection{Stellar rotation period and differential rotation}

The TESS data show a large amplitude of the light curve in the $I_{\rm c}$ band ($0.3207\pm0.0006$~mag), reflecting the strong activity level of LkCa~4, that remains stable over the 49~d of monitoring. We modeled the very accurate TESS data using a quasi-periodic GP to refine the stellar rotation period, found to be equal to $3.372\pm0.002$~d over our observing window, compatible with a previous estimate (3.374; \citealt{grankin08}). Other indicators derived from SPIRou data, such as the longitudinal field and the RVs, also yield consistent stellar rotation periods. More specifically, the GPR fit to the RVs provide the same value of $3.372\pm0.002$~d while the longitudinal field measurements suggest a slightly larger value of $3.373\pm0.002$~d, though still compatible.

We also find that the surface of LkCa~4 is sheared by a weak level of DR, $9.5\pm1.3$ times weaker than that of the Sun. In addition, we note that the estimate of the surface DR is much more constrained from our Stokes~$V$ LSD profiles, with error bars about 3 times smaller than those derived from Stokes~$I$ LSD profiles.
Although estimates from optical data are consistent with ours, the error bars derived from SPIRou data (from Stokes~$I$ and $V$) are smaller than those obtained by \cite{donati14}, up to a factor 30 for Stokes~$V$ profiles, which can be largely explained by our larger number of observations (41 with SPIRou vs. 12 with ESPaDOnS) spread over a larger time interval ($108$~d with SPIRou vs. 13~d with ESPaDOnS) and the benefits of the enhanced Zeeman effect in the NIR (for DR estimates from Stokes~$V$ data).
We note that our new results indicate that the DR is significantly different from 0 (solid-body rotation) unlike those of \cite{donati14}.

\subsection{Spot coverage}

From our brightness reconstructions with ZDI, we found that the stellar surface is covered with spots/plages at a level of about 7\% when considering SPIRou data alone, or 9\% when including TESS photometry in the fitting process. These values are much lower than those derived from optical data ($\unsim25$\%; \citealt{donati14}), in agreement with the expected decrease in the brightness contrast of surface features with increasing wavelength. In addition, we do not see a clear polar spot as in ESPaDOnS data, but rather an elongated spot spreading from the pole to the equator around phase 0.1, that could simply result from an evolution of the brightness distribution between 2014 and 2022.

We note that adding TESS photometry in the fitting process increases the spot coverage by $\unsim2$\%, mainly by enhancing the contrasts of brightness features and adding low-latitude structures to ensure that both spectroscopic and photometric data are fitted down to a unit $\chi^2_r$. This increase is similar to what has been observed for V410~Tau \citep{finociety21} and most likely reflects that spectroscopic and photometric data are not sensitive to the same surface structures. 

Our Stokes~$I$ LSD profiles show a clear modulation of their EWs with rotation phase (up to 20\%), being minimum at phase 0.1 (i.e. when the cool spot is visible) and maximum at phase 0.75 (i.e. when the warm plage is visible). To take into account these variations during the fitting process with ZDI, the depth of the local profiles is allowed to vary as a power $\delta$ of the local brightness. Assuming a constant EW (i.e. $\delta=0$) as in all previous ZDI studies, yields a more contrasted brightness map (spot coverage of $\unsim15$\%) the associated light curve of which in the TESS bandpass has an amplitude twice larger than the observed one. We therefore empirically find that, for this star, $\delta=1.0\pm0.1$ allows us to reproduce the amplitude of the EWs variations as well as the amplitude of the observed light curve. Most of previous ZDI studies dealt with optical data for which variations in EWs are not significant.  

\begin{figure}
    \centering
    \includegraphics[scale=0.3,trim={1cm 0cm 1cm 1cm},clip]{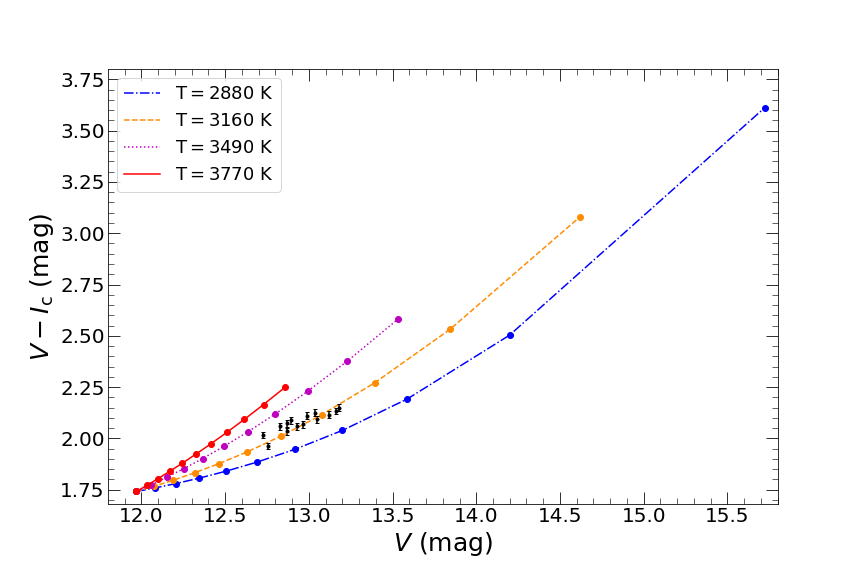}
    \caption{$V-I_{\rm c}$ colour indexes as a function of the magnitude in the $V$ band, collected from the ground-based AZT-11 telescope at CrAO between 2021 Oct and 2022 Jan. The black dots represent the measurements with the empirical error bars as derived in Sec.~\ref{sec:ground_based_observations}. Each colour line represent a two-temperature model with a fixed photospheric temperature of 4140~K and a fixed spot temperature of 2880 (blue dash-dotted line), 3160 (orange dashed line), 3490 (magenta dotted line) and 3770~K (red solid line). Each filled circle corresponds to a different filling factor for the spot with steps of 10\%. The unspotted magnitude therefore corresponds to $V=11.97$. Our data are consistent with spots at 3160~K covering about 70\% of the stellar surface. }
    \label{fig:two_temperature_model}
\end{figure}

ZDI is mostly sensitive to large structures at the surface of the star and misses most of the small features. To estimate the percentage of the visible surface of LkCa~4 covered by dark spots, we used a two-temperature model to fit our $V-I_{\rm c}$ indexes as a function of the $V$ magnitude, taking into account the visual extinction $A_{\rm V}$ derived in \cite{donati14} and the colour indexes for young stars from \cite{pecaut13}. We assumed a fixed photospheric temperature of 4140~K and a fixed temperature for the spots, with various filling factors. We found that the $V-I_{\rm c}$ indexes are consistent with about 70\% of the visible surface being covered by spots at 3160~K (Fig.~\ref{fig:two_temperature_model}) if assuming a unspotted magnitude of 11.97 (as derived from the temperature, distance, radius and visual extinction). The high spot coverage is consistent with the estimate of \cite{gullysantiago17} and similar to what is observed for V410~Tau \citep{yu19,finociety21}. However, \cite{gullysantiago17} reported a spot temperature between 2700 and 3000~K, thus lower than our estimate. Our model does not take into account the bright plages which may lead to an overestimate of the temperature of cool spots. 

\subsection{Magnetic topology}

The typical amplitude of our Stokes~$V$ LSD profiles (0.5\% of the unpolarized continuum) is typically twice smaller than that of optical data collected with ESPaDOnS (1\% of the unpolarized continuum, \citealt{donati14}). However, NIR observations, being more sensitive to the magnetic field, provide more accurate measurements than optical data in about half the exposure time. In particular, the larger noise level in our Stokes~$V$ LSD profiles ($3.4\times10^{-4}$) corresponds to the typical noise level in the previous optical study ($3.5\times10^{-4}$; \citealt{donati14}). 

We were therefore able to reliably reconstruct the magnetic topology from our set of Stokes~$I$ and $V$ LSD profiles. We find that the magnetic topology of LkCa~4 is rather simple and very similar to the one derived from optical data \citep{donati14}. The poloidal field encloses about 65\% of the magnetic energy, consistent with previous estimates. From our data, we estimate that the poloidal field mainly consists of a dipole (of polar strength $\unsim2.2$~kG, with an axis tilted by $\unsim3^\circ$ with respect to the rotation axis, towards phase 0.31) that concentrates nearly 85\% of the poloidal field energy. This dipole is about 35\% stronger than the one reconstructed in \cite{donati14} and is better aligned with respect to the rotation axis of the star (from $10^\circ$ in 2014 to $3^\circ$ in 2021-2022). 
The toroidal component consists of an equatorial ring of intense azimuthal field, reaching strengths of $\unsim1.4$~kG. These results suggest that the magnetic topology remained more or less stable over several years with a slight increase in strength. Future observations of LkCa~4 would help to clarify whether the magnetic field of LkCa~4 does significantly evolve with time, e.g., switching overall polarity at times, or is rather in a more or less steady-state dynamo regime.

\subsection{Radial velocities}

From our Stokes~$I$ LSD profiles, we derived RVs of LkCa~4 that show a full amplitude of 6.10~\kms\ with a dispersion of 1.45~\kms. The observed full amplitude is $1.4\times$ larger than the one derived from ESPaDOnS data \citep{donati14}. This most likely reflects that the brightness distribution at the surface of LkCa~4 has significantly evolved between 2014 and 2022, generating now a stronger RV activity jitter. 

In addition, our typical error bars derived from photon-noise (0.15~\kms) are $2.5\times$ larger than those derived from ESPaDOnS data (0.06~\kms; \citealt{donati14}).

Fitting our RV measurements with a quasi-periodic GP yields an excess of uncorrelated noise of $s=0.37$~\kms, indicating that the measurements are also affected by stochastic noise (most likely due to stellar activity) that is not taken into account when estimating the error bars from photon noise only.

We first modeled the measured RVs from the ZDI brightness reconstructions. 
Accounting for TESS photometry in ZDI slightly degrades the filtering of the activity jitter as the RMS dispersion of the filtered RVs increases by a factor $\unsim1.07$ with respect to the model including spectroscopic data only (from 0.45~\kms\ to 0.48~\kms). This effect is not as strong as in the case of V410~Tau for which the RMS dispersion of the filtered RVs varies by a factor $\unsim1.8$ between both ZDI models \citep{finociety21}. The model associated with the quasi-periodic GP yields a slightly lower RMS dispersion of the filtered RVs of 0.38~\kms. 
The RMS dispersions of the filtered RVs obtained from our ZDI models and GPR are about 12 to 16 times smaller than the full amplitude of the measured RVs. The filtering achieved by \cite{donati14} using the same method on optical data was better, as the ratio between the full amplitude and the dispersion of their filtered RVs reached a value of 78. The difference most likely reflects the changes in brightness distribution between both epochs, and presumably an increase in the amount of intrinsic variability as well.

We see no significant modulation of all our filtered RVs that would indicate the presence of a close-in massive planet, in agreement with \cite{donati14}. From simulated data, we estimate that our observations would allow one to detect a planet if the semi-amplitude of its RV signature is larger than 0.44~\kms\ and its period is shorter than 50~d. This value is larger than the first estimate by \cite{donati14} as the noise level in our data (due to both photon-noise and intrinsic variability) is almost 7$\times$ larger. Our threshold typically corresponds to a 4.3~M$_{\rm jup}$ planet at a distance of 0.1~au ($P=13$~d) or a 6.3~M$_{\rm jup}$ planet at a distance of 0.2~au ($P=36.8$~d). These upper limits are mainly limited by the high level of intrinsic variability due to stellar activity during the time-span of our data set. More observations would be needed to further assess the potential presence of a close-in massive planet.

\subsection{Chromospheric activity}

The \ion{He}{i} triplet at 1083~nm, the Pa$\beta$ and Br$\gamma$ lines were used as proxies of the chromospheric activity of LkCa~4. We first computed the EWVs of the three lines. The \ion{He}{i} triplet shows a modulation of its EWVs with rotational cycle (with a probability that this modulation is spurious by chance of 0.15) for which enhanced absorption takes place slightly before the visible pole of the dipole crosses the line-of-sight (phase 0.4) and cannot be related to any magnetic feature. The Pa$\beta$ EWVs exhibit a more significant modulation for which enhanced absorption (at phase 0.3) may relate to the absorption by wind material escaping the star along open field lines when the pole of the dipole faces the observer. 
The 2D periodograms further confirms the modulation of the \ion{He}{i} and Pa$\beta$ lines, with in particular half the rotation period showing up in the Pa$\beta$ 2D periodogram. The \ion{He}{i} and Pa$\beta$ autocorrelation matrices also indicate that an activity signal is detected across the line width. 

These results suggest that these two lines react differently to the magnetic field and stellar activity since the modulation of the Pa$\beta$ line is more significant than that of the \ion{He}{i} triplet in contrast to what has been observed for the similar wTTS V410~Tau \citep{finociety21} and the phased EWV curves of both lines do not match well. However, we find no evidence of a rotationally modulated activity signal in the Br$\gamma$ line as for V410~Tau \citep{finociety21}. More observations with NIR instruments are needed to investigate whether some trends can be highlighted in the behavior of these three lines in active wTTSs.

\subsection{Conclusion}

As a follow-up analysis to that of \cite{donati14}, our study confirms most of the previous results from a completely new data set and wavelength domain. New observations of LkCa~4 are needed to further constrain the evolution of the large-scale magnetic field of this star (e.g. magnetic cycle or steady-state dynamo regime). Observations of other wTTSs with SPIRou, including those targeted within the SLS, will allow one to investigate in details the role that magnetic field plays in the stellar and planetary formation during the transition phase between cTTSs and MS stars. Contemporaneous optical and NIR spectropolarimetric observations of PMS stars would be extremely useful to improve the activity-modelling and jitter-filtering methods, as both domains provide complementary information on stellar activity. Such capabilities would be a key asset for the study of the stellar activity and its impact on the RV data to better characterize planetary systems around active PMS stars like AU~Mic \citep{plavchan20,klein21,klein22} and V1298~Tau \citep{david19a,david19b,suarez-mascareno21}.  

%%%%%%%%%%%%%%%%%%%%%%%%%%%%%%%%%%%%%%%%%%%%%%%%%%
\section*{Acknowledgements}

This work includes data collected in the framework of the SPIRou Legacy Survey (SLS), an international large programme that was allocated on the Canada-France-Hawaii Telescope (CFHT) at the summit of Maunakea by the Institut National des Sciences de l’Univers of the Centre National de la Recherche Scientifique of France, the National Research Council of Canada, and the University of Hawaii. We acknowledge funding by the European Research Council (ERC) under the H2020 research \& innovation programme (grant agreements \#740651 NewWorlds, \#716155 SACCRED, \#743029 EASY).

\section*{Data Availability}

The data collected with TESS are publicly available from the Mikulski Archive for Space Telescopes (MAST) website.
The SPIRou data collected as part of the SLS will be publicly available from the CADC website one year after the completion of the SLS programme, i.e. by mid-2023.



\bibliographystyle{mnras}
\bibliography{main} % if your bibtex file is called example.bib

%\begin{thebibliography}{99}
%\end{thebibliography}

%%%%%%%%%%%%%%%%%%%%%%%%%%%%%%%%%%%%%%%%%%%%%%%%%%

%%%%%%%%%%%%%%%%% APPENDICES %%%%%%%%%%%%%%%%%%%%%

\appendix

\section{Journal of observations for ground-based photometry}

We provide a full journal for the observations collected with the ground-based AZT-11 telescope at CrAO in Table~\ref{tab:log_crao}.

\begin{table}

\caption{Ground-based photometric observations of LkCa~4 collected with the AZT-11 telescope at CrAO between 2021 October and 2022 January. The $1^{\rm st}$ and $2^{\rm nd}$ columns list the date and the Heliocentric Julian Date. In column 3, we give the measured magnitude in the $V$ band. Columns 4 and 5 report the colour indexes $V-R_{\rm J}$, $V-I_{\rm J}$ in the Johnson system while columns 6 and 7 detail the colour indexes $V-R_{\rm c}$ and $V-I_{\rm c}$ in the Cousins system.}
\label{tab:log_crao}
\centering 
\resizebox{0.49\textwidth}{!}{
\begin{tabular}{lcccccc}
\\
\hline \hline
\multicolumn{1}{c}{Date} & HJD  & $V$ & $V-R_{\rm J}$ & $V-I_{\rm J}$ & $V-R_{\rm c}$ & $V-I_{\rm c}$ \\ 
 &  2459000+ & (mag) & (mag) & (mag) & (mag) & (mag)  \\ \hline

2021 October 10 & 498.417 & 12.932 & 1.515 & 2.624 & 1.053 & 2.059 \\ 
2021 October 17 & 505.509 & 13.121 & 1.498 & 2.696 & 1.041 & 2.115 \\ 
2021 October 30 & 518.519 & 12.868 & 1.556 & 2.594 & 1.083 & 2.035 \\ 
2021 November 03 & 522.524 & 13.181 & 1.575 & 2.740 & 1.096 & 2.150 \\ 
2021 November 10 & 529.514 & 13.160 & 1.560 & 2.718 & 1.086 & 2.133 \\ 
2021 November 11 & 530.521 & 12.987 & 1.559 & 2.689 & 1.085 & 2.110 \\ 
2021 November 12 & 531.535 & 12.728 & 1.539 & 2.569 & 1.071 & 2.016 \\ 
2021 November 15 & 534.519 & 12.830 & 1.578 & 2.622 & 1.099 & 2.058 \\ 
2021 November 16 & 535.535 & 12.963 & 1.520 & 2.637 & 1.057 & 2.069 \\ 
2021 December 02 & 551.219 & 12.891 & 1.570 & 2.662 & 1.093 & 2.088 \\ 
2021 December 10 & 559.187 & 13.037 & 1.580 & 2.708 & 1.100 & 2.125 \\ 
2021 December 23 & 572.321 & 12.758 & 1.475 & 2.501 & 1.025 & 1.963 \\ 
2022 January 25 & 605.247 & 12.871 & 1.568 & 2.641 & 1.091 & 2.073 \\ 
2022 January 27 & 607.306 & 13.050 & 1.543 & 2.667 & 1.074 & 2.093 \\ \hline

\end{tabular} }
\end{table}

\section{\ion{He}{i}, Pa$\beta$ and Br$\gamma$ spectra}

\begin{figure}
    \centering \hspace*{-1.1cm}
    \begin{subfigure}[b]{0.49\textwidth}
         \centering
         \includegraphics[scale=0.33]{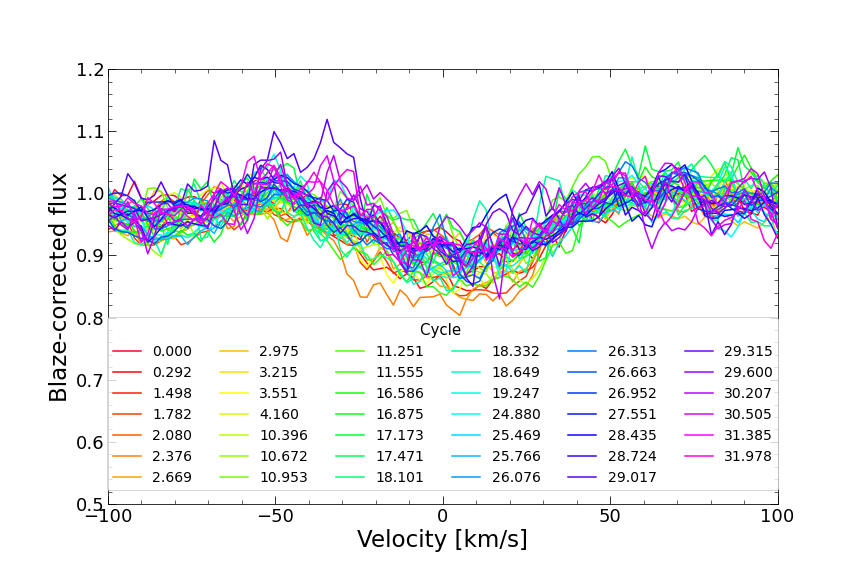}
         
    \end{subfigure}

    \hspace*{-1.1cm}
    \begin{subfigure}[b]{0.49\textwidth}
         \centering
         \includegraphics[scale=0.33]{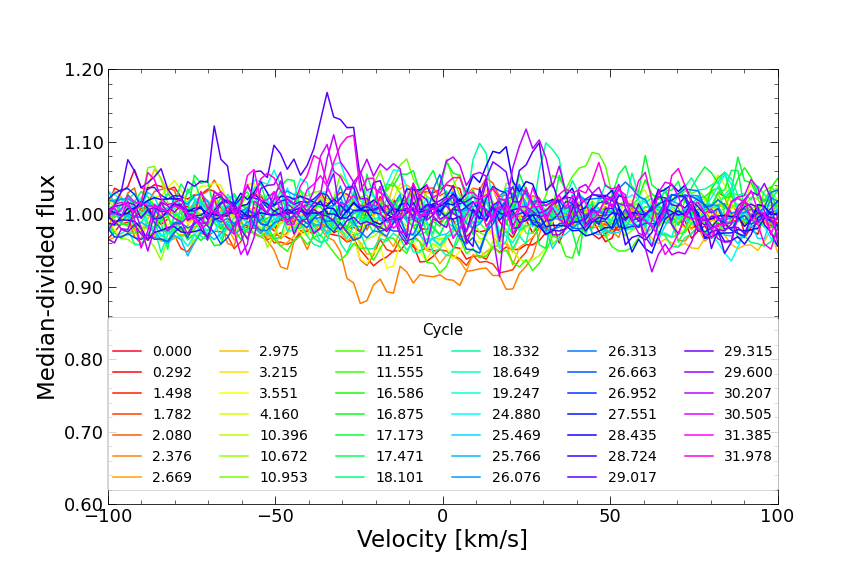}
         
    \hspace*{-4.5cm}
    \begin{subfigure}[b]{0.49\textwidth}
         \centering
         \includegraphics[scale=0.33]{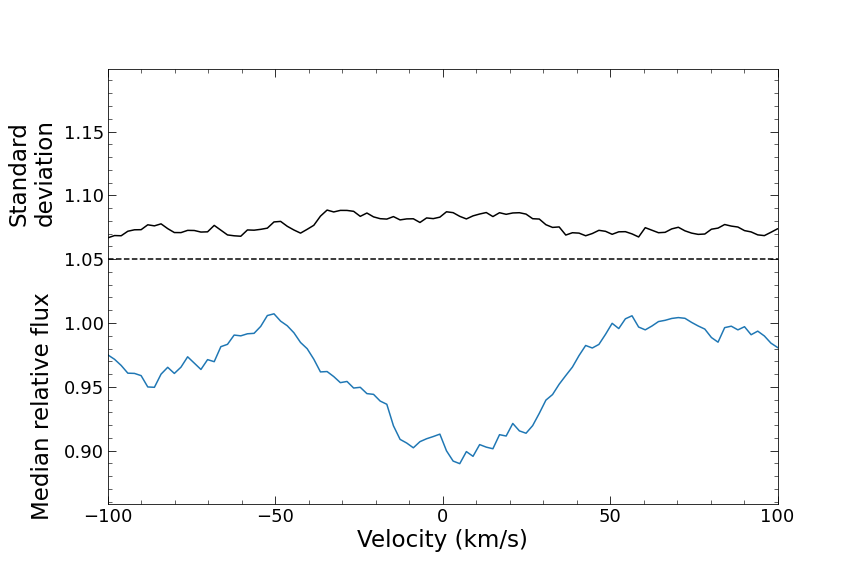}
       
    \end{subfigure}
         
    \end{subfigure}
    \caption{Telluric corrected profiles (top), median-divided profiles (middle) and median profile (bottom) for the \ion{He}{i} triplet. In the bottom panel, the dispersion in the velocity bins of the median-divided spectra is shown in solid black line, shifted upwards by 1.05 for display purposes while the dashed line depicts the zero variability level.} 
    \label{fig:spectral_line_he}
\end{figure}

\begin{figure}
    \centering \hspace*{-1.1cm}
    \begin{subfigure}[b]{0.49\textwidth}
         \centering
         \includegraphics[scale=0.33]{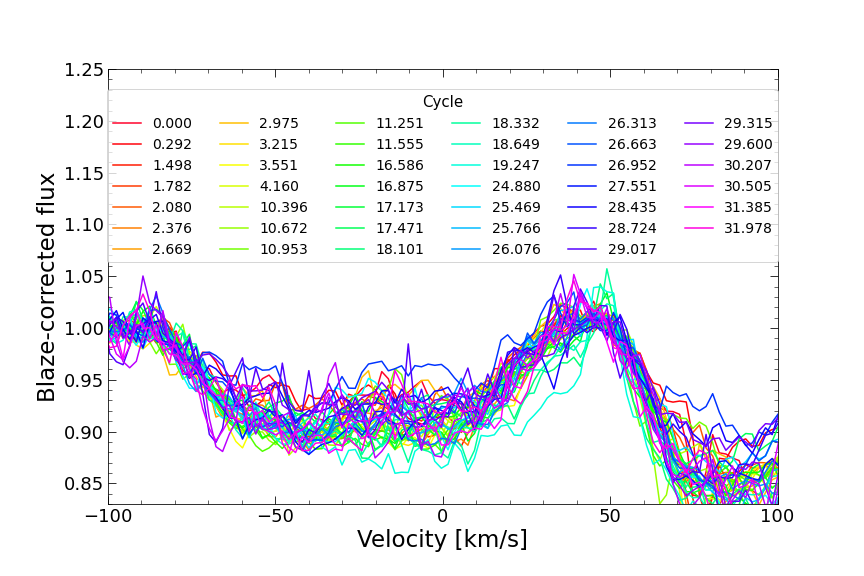}
         
    \end{subfigure}

    \hspace*{-1.1cm}
    \begin{subfigure}[b]{0.49\textwidth}
         \centering
         \includegraphics[scale=0.33]{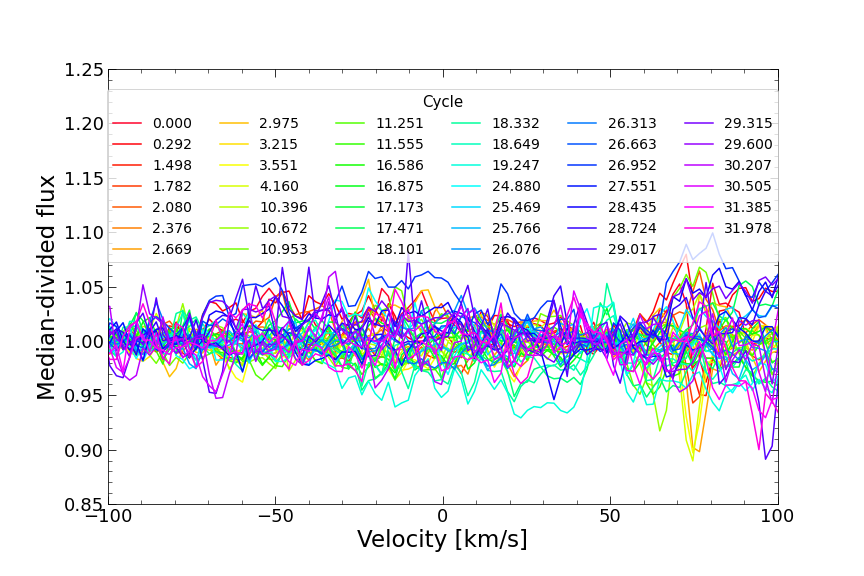}
         
    \hspace*{-4.5cm}
    \begin{subfigure}[b]{0.49\textwidth}
         \centering
         \includegraphics[scale=0.33]{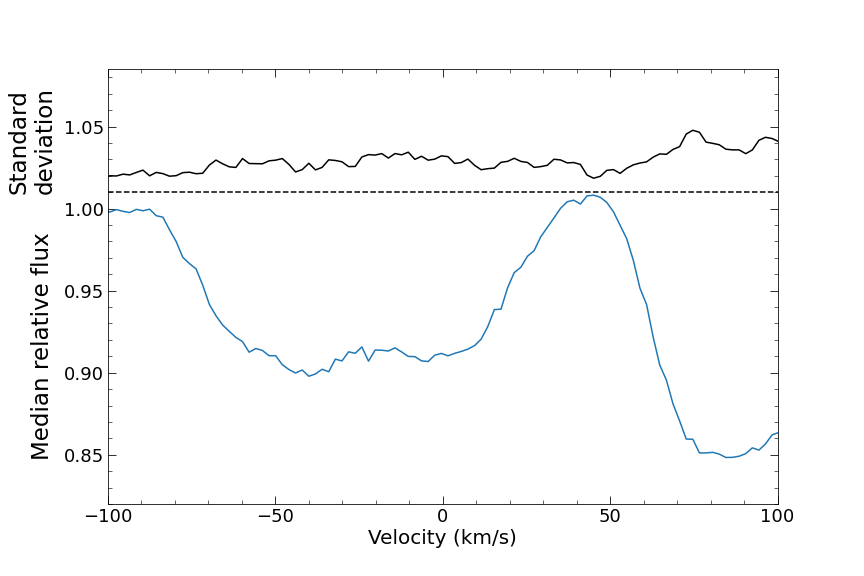}
       
    \end{subfigure}
         
    \end{subfigure}
    \caption{Same as Fig.~\ref{fig:spectral_line_he} for the Pa$\beta$ line. In the bottom panel, the zero variability level is shifted upwards by 1.01 for clarity purposes. In addition, we see that the blue wing of the line is blended, likely with a Ca line, causing the depression around -50~\kms.} 
    \label{fig:spectral_line_pa}
\end{figure}

\begin{figure}
    \centering \hspace*{-1.1cm}
    \begin{subfigure}[b]{0.49\textwidth}
         \centering
         \includegraphics[scale=0.33]{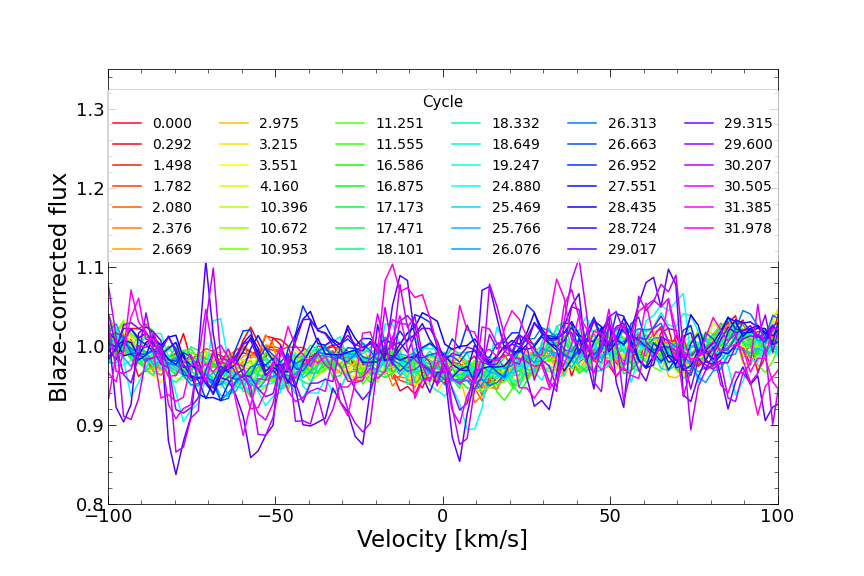}
         
    \end{subfigure}

    \hspace*{-1.1cm}
    \begin{subfigure}[b]{0.49\textwidth}
         \centering
         \includegraphics[scale=0.33]{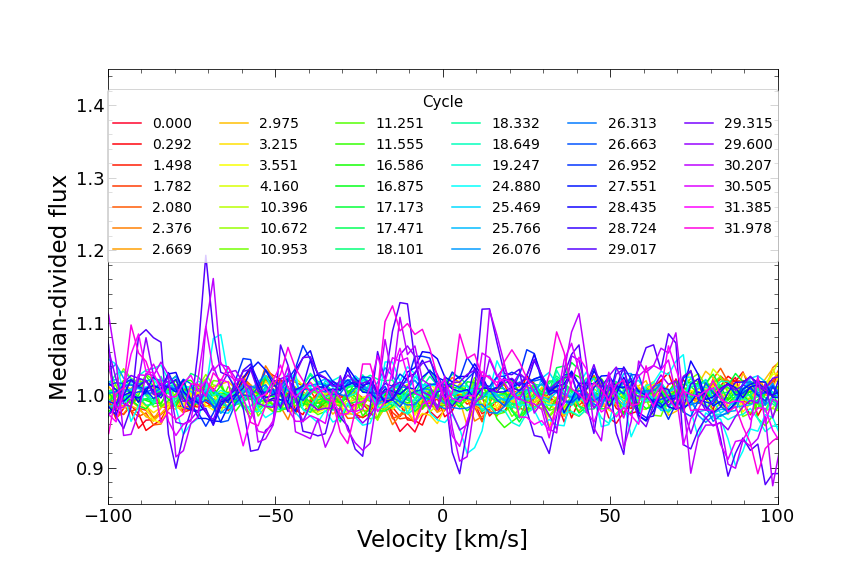}
         
    \hspace*{-4.5cm}
    \begin{subfigure}[b]{0.49\textwidth}
         \centering
         \includegraphics[scale=0.33]{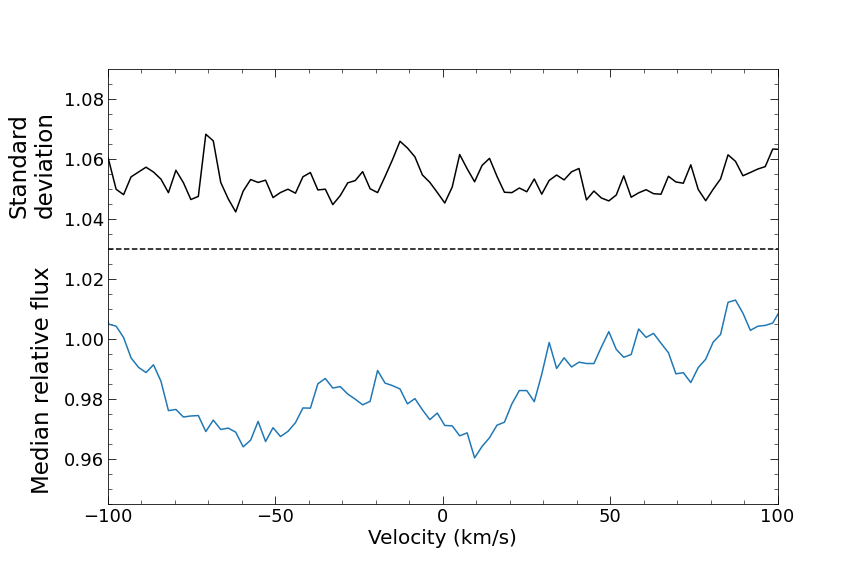}
       
    \end{subfigure}
         
    \end{subfigure}
    \caption{Same as Fig.~\ref{fig:spectral_line_he} for the Br$\beta$ line. In the bottom panel, the zero variability level is shifted upwards by 1.03 for clarity purposes. We see essentially noise in this line.} 
    \label{fig:spectral_line_br}
\end{figure}

We show the telluric-corrected spectra, median spectrum and median-divided spectra for the \ion{He}{i} (Fig.~\ref{fig:spectral_line_he}), Pa$\beta$ (Fig.~\ref{fig:spectral_line_pa}) and Br$\gamma$ (Fig.~\ref{fig:spectral_line_br}) lines, used as proxies to investigate the chromospheric activity of LkCa~4.

%%%%%%%%%%%%%%%%%%%%%%%%%%%%%%%%%%%%%%%%%%%%%%%%%%

% Don't change these lines
\bsp	% typesetting comment
\label{lastpage}
\end{document}